\documentclass[10pt]{article}

\usepackage{amscd,amssymb,amsmath,amsfonts,bm}
\usepackage[mathcal,mathscr]{euscript}
\usepackage{graphicx}
\usepackage[framemethod=tikz]{mdframed}
\usepackage{bigfoot}
\usepackage{float}

\usepackage{color}
\usepackage{listings}

\newcommand{\ind}{\mathrm{ind}}
\newcommand{\set}[1]{\left\lbrace{#1}\right\rbrace}
\newcommand{\red}{\color{red}}
\newcommand{\blue}{\color{blue}}

\usepackage{hyperref}
\usepackage{float}
\usepackage{amscd,amssymb,amsmath,latexsym,bm}
\usepackage[mathcal,mathscr]{euscript}
\usepackage{lipsum}
\usepackage{amsfonts}
\usepackage{graphicx}
\usepackage{epstopdf}
\usepackage{algorithmic}
\usepackage{hyperref}
\usepackage{xcolor}
\usepackage{upgreek}
\usepackage{enumerate}
\usepackage{ulem}
\usepackage{stmaryrd}
\usepackage{gensymb}			%DJA added this for degrees symbol
\usepackage[utf8]{inputenc}

\numberwithin{equation}{section}
\numberwithin{figure}{section}

\newcommand{\CM}{{\mathbb C}}

\newcommand{\ZM}{{\mathbb Z}}

\newcommand{\Ee}{{\mathcal E}}
\newcommand{\Ii}{{\mathcal I}}

\newcommand{\Vv}{{\mathcal V}}

\newcommand{\Tt}{{\mathcal T}}

\newcommand{\Mm}{{\mathcal M}}

\newcommand{\Jj}{{\mathcal J}}

\newcommand{\Ll}{{\mathcal L}}

\newcommand{\Kk}{{\mathcal K}}
\newcommand{\Hh}{{\mathcal H}}

\begin{document}

%\title{A Computer Code for Topological Quantum Spin Systems over Triangulated Surfaces}

%\author{Yingkai Liu and Emil Prodan}

%\address{Department of Physics and Department of Mathematical Sciences 
%\\Yeshiva University, New York, NY 10016, USA \\
%\href{mailto:prodan@yu.edu}{prodan@yu.edu}}

\title{A Computer Code for Topological Quantum Spin Systems over Triangulated Surfaces}

\author{Yingkai Liu and Emil Prodan\\
\small Department of Physics, Yeshiva University\\
\small New York, New York 10016, USA}

%\author{Emil Prodan}

%\address{Department of Physics, Yeshiva University, Address\\
%New York, New York 10016, USA\\
%prodan@yu.edu}

\date{}

\maketitle

\begin{abstract} 
We derive explicit closed-form matrix representations of Hamiltonians drawn from tensored algebras, such as quantum spin Hamiltonians. These formulas enable us to soft-code generic Hamiltonian systems and to systematize the input data for uniformly structured as well as for un-structured Hamiltonians. The result is an optimal computer code that can be used as a black box that takes in certain input files and returns spectral information about the Hamiltonian. The code is tested on Kitaev's toric model deployed on triangulated surfaces of genus 0 and 1. The efficiency of our code enables these simulations to be performed on an ordinary laptop. The input file corresponding to the minimal triangulation of genus 2 is also supplied.
\end{abstract}

{\scriptsize \tableofcontents}

\setcounter{tocdepth}{1}

\section{Introduction}

The research efforts towards a multi-purpose quantum computer have accelerated in the past years on both the hardware and software fronts \cite{GibneyNature2019}. A major event in the field was Kitaev's proposal and realization of a theoretical fault-tolerant quantum computation platform based on braiding topological defects of topological phases of matter \cite{KitaevAOP2003,KitaevAOP2006}. For example, the principles behind the quantum computer based on Majorana anyons in superconducting wires \cite{SarmaQuantum2015} fall within this framework. The advantages of a topological qubit have been already probed experimentally in \cite{NiggScience2014} using a trapped-ions platform. The principles of topological computation, however, are certainly not easy to comprehend and these extremely interesting ideas are vehiculated and evolved within a rather small community of experts. Indeed, the theoretical models supporting topological quantum defects consist of quantum double algebras deployed on triangulations and networks, which result in extremely complex Hamiltonians belonging to tensored algebras. One such Hamiltonian, the classic toric code introduced by Kitaev \cite{KitaevAOP2003}, is presented in Section~\ref{Sec:Applications}. It can be deployed on many triangulations or it can be perturbed with local and string-like operators and the quantum algorithms depend crucially on the spectral properties of not just one Hamiltonian but of a whole family of Hamiltonians. Having access to and control over this spectral information is crucial for both the software and hardware development and, in authors' opinion, developing computer codes that deliver just that could aid not only the research in the field but also the passing of the know-how to younger generations and to engineers. Such numerical tools become accessible to a large community if they fulfill at least two criteria: 1) The user only needs to prepare input files when new applications are considered; 2) The codes can be executed with modest computational resources, preferably with an ordinary laptop. Our work has been motivated and shaped by these two targets.

\vspace{0.2cm}

We will be dealing exclusively with the algebra $\Mm_m^{\otimes L}$ of $L$-fold tensor product of the elementary algebra $\Mm_m$ of $m\times m$ complex matrices. One strategy for building a generic computer code for this tensored algebra, which can deal with arbitrary Hamiltonians or large classes of Hamiltonians, is to rely entirely on the action of operators of the type $I\otimes \ldots I \otimes A \otimes I \ldots \otimes I$ that generate the tensored algebra. Indeed, since any other operator can be generated as products and sums of such operators, a generic operator can be coded as a repeated action of generating operators. Unfortunately, this results in highly inefficient code. The alternative is to specialize the code, {\it i.e.} to hard-code the action of higher-rank products such as $I\otimes \ldots I \otimes A_1 \otimes A_2 \otimes I \ldots \otimes I$, which appear quite often in spin-chain Hamiltonians. While such an approach works well for linear spin-chains, for spin systems over triangulations, the local environment, the spin couplings and even the rank of the tensor products change from one triangulation to another. This means a scientist will have to practically re-code the entire script every time when the triangulation is modified.

\vspace{0.2cm}

The contribution reported in this paper provides numerical tools as well as complete and working computer codes that can be used as black boxes, taking in certain input files and returning spectral information about the Hamiltonians. In other words, we found a way to simulate large classes of Hamiltonians without modifying the core of the source code. When the Hamiltonians are structured as in the case of the unperturbed Kitaev model, the information contained in the input files can be extracted directly from the triangulation and, as such, the task of generating input files can be completed entirely independent. Our approach relies on closed-form matrix expressions for operators of the form:
	\begin{equation}\label{Eq:Op1}
	I\otimes \ldots \otimes I \otimes A^{1} \otimes
	I \otimes \cdots  \otimes
	 I\otimes A^{Q-1} \otimes
	I \otimes \cdots \otimes I , \quad A^i, I \in \Mm_m,
	\end{equation}
The search for such expressions was prompted by a similar development \cite{ProdanArxiv2019} in the context of the algebra generated by fermionic creation and annihilation operators. These expressions enable us to:
\begin{itemize}
\item Populate the $m^L \times m^L$ matrices corresponding to the operators \eqref{Eq:Op1} in a minimal number of steps, hence generate optimal computer codes. 
\item Write one generic block of code that covers all operators \eqref{Eq:Op1}, regardless of their particular structure.
\item Structure the input files to a standard and transparent format.
\end{itemize}

\vspace{0.2cm}

The paper is structured as follows. Section~\ref{Sec:ElemOp} presents the derivation of the closed-form matrix representations of operators of type \eqref{Eq:Op1}. This section also shows how to structure the data related to \eqref{Eq:Op1} into well-defined functions, which later will decide the structure of the computer codes. The reader will also find here our solution for soft-coding the expressions, which rely on encoding multi-indices into linear global indices using generalized number bases. Section~\ref{Sec:SpinH} introduces the spin Hamiltonians over triangulations. The terms appearing in these Hamiltonians are of the form \eqref{Eq:Op1} but the indices $1,\ldots,Q-1$ are now associated with subsets of the triangulation. The reader will also find here detailed presentations of the main blocks of our computer codes. Section~\ref{Sec:Applications} contains our applications consisting of full and partial diagonalizations of the standard toric model on triangulations of genus 0 and 1. The reader will find here the input files associated with those triangulations as well as a triangulation of genus 2. Appendix~\ref{App:1} supplies a complete working Fortran code that computes and diagonalizes the full Hamiltonian. Appendix~\ref{App:2} and Appendix~\ref{App:3} supply a Fortran codes that resolve the low energy spectrum of the Hamiltonian. The last section supply facts about the performance of our code when executed on a Microsoft Surface Laptop 2 with different input files.
	
\section{Elementary Operators}
\label{Sec:ElemOp}
	
In this section we will be dealing with local Hilbert spaces $\CM^m$ and local algebras that are canonically isomorphic to that of $m \times m$ matrices with complex entries, denoted here by $\Mm_m$. The unit element of $\Mm_m$ will be denoted by $I_m$. We will also be dealing with the global Hilbert space ${\left(\mathbb C^m\right)}^{\otimes L}$ and the global algebra of observables $\Mm_m^{\otimes L}$. Our focus will be on elementary operators of the type:
	\begin{equation}\label{Eq:ElementaryOp}
	I_m^{\otimes {n_1}} \otimes A^{1} \otimes
	I_m^{\otimes {n_2}} \otimes
	\cdots\otimes A^{Q-1} \otimes
	I_m^{\otimes {n_{Q}}} , \quad A^i \in \Mm_m,
	\end{equation}
acting on the global Hilbert space. The indices in \eqref{Eq:ElementaryOp} are assumed to obey the constraint  $L=Q-1+\sum_{s=1}^{Q} n_Q$. The goal of this section is to derive a closed-form matrix representation of \eqref{Eq:ElementaryOp}, for an arbitrary input set of data $\{Q,m,\bm n, \bm A\}$. Throughout, sequences and multi-indices will be compactified in a corresponding bold letter, such as $\bm n$ for $\{n_1,n_2,\ldots\}$. 
	
\subsection{Basic Relations}
	
The elementary matrix from $\Mm_\alpha$ with one single non-zero entry of value 1 at position $(i,j)$ will be denoted by $E^{(\alpha)}_{i,j}$.	Our matrix representations will be derived by applying and iterating the following basic facts: 
	\begin{itemize}
	\item Any matrix $A \in \Mm_\alpha$ can be decomposed as a sum of elementary matrices:
	\begin{equation}
	A = \sum_{i,j=1}^\alpha A_{i,j} \, E^{(\alpha)}_{i,j}.
	\end{equation} 
	\item 	The rule for the tensor product of two elementary matrices is:
		\begin{equation}
		E^{(\alpha)}_{i,j} \otimes E^{(\beta)}_{k,l} = E^{(\alpha \cdot \beta)}_{\beta(i -1)+k, \beta (j-1) +l}
	    \end{equation}	
	    \item The tensor product distributes over addition:
		\begin{equation}
		(x+y)\otimes (z+w) = x\otimes z+x\otimes w+y\otimes z + y \otimes w.
		\end{equation}
	\end{itemize}
	
Obviously $\Mm_m^{\otimes L} \simeq \Mm_{m^L}$, hence the operators \eqref{Eq:ElementaryOp} accept a decomposition of the form:
\begin{equation}
\sum_{i,j=1}^{m^L} M_{i,j} \, E^{(m^L)}_{i,j}.
\end{equation}
Then our task is to identify the complex coefficients $M_{i,j}$.
	
\subsection{Patterns in Lower Rank Tensor Products}	
	
We derive the general matrix expression of \eqref{Eq:ElementaryOp} using an inductive process. Inherently, during our computations, the indices will develop complex expressions and,	when these expressions are too long to be placed on the same line, the indices will be specified as $E^{(\alpha)}_{\scriptsize\set{\substack{\cdots \\ \cdots}}}$. We present first the incipient steps of the induction argument, namely, the computation of a few lower rank tensor products:
		\begin{align}
		I_m^{\otimes {n_1}} \otimes A^{1} 
		=		\sum_{i_1=1}^{m^{n_1}}\sum_{j_1,j'_1=1}^{m} A_{j_1,j'_1} E^{(m^{n_1+1})}_{m(i_1-1)+j_1,m(i_1-1)+j'_1}
		\end{align}
	\begin{align}\label{Eq:T3}
		I_m^{\otimes {n_1}} \otimes A^{1} \otimes I_m^{\otimes {n_2}} =		\sum_{i_1=1}^{m^{n_1}}\sum_{i_2=1}^{m^{n_2}} \sum_{j_1,j'_1=1}^{m} A_{j_1,j'_1} E^{(m^{n_1+n_2+1})}_{\scriptsize\set{\substack{m^{n_2}(m(i_1-1)+j_1-1)+i_2\\ m^{n_2}(m(i_1-1)+j'_1-1)+i_2}}}
\end{align}
\begin{align}
		& I_m^{\otimes {n_1}} \otimes A^{1} \otimes I_m^{\otimes {n_2}} \otimes A^{2} = \\ \nonumber
		& \qquad 		\sum_{i_1=1}^{m^{n_1}}\sum_{i_2=1}^{m^{n_2}} \sum_{j_1,j'_1=1}^{m}\sum_{j_2,j'_2=1}^{m} A_{j_1,j'_1}A_{j_2,j'_2} E^{(m^{n_1+1+n_2+1})}_{\scriptsize\set{\substack{m(m^{n_2}(m(i_1-1)+j_1-1)+i_2-1)+j_2\\ m(m^{n_2}(m(i_1-1)+j'_1-1)+i_2-1)+j'_2}}}
		\end{align}
		\begin{align}\label{Eq:T5}
		&I_m^{\otimes {n_1}} \otimes A^{1} \otimes I_m^{\otimes {n_2}} \otimes A^{2} \otimes I_m^{\otimes {n_3}} = \\ \nonumber
		& \qquad 	
		\sum_{i_1=1}^{m^{n_1}}\sum_{i_2=1}^{m^{n_2}}\sum_{i_3=1}^{m^{n_3}} \sum_{j_1,j'_1=1}^{m}\sum_{j_2,j'_2=1}^{m} A_{j_1,j'_1}A_{j_2,j'_2} 
		 E^{(m^{n_1+1+n_2+1+n_3})}_{\scriptsize\set{\substack{m^{n_3}(m(m^{n_2}(m(i_1-1)+j_1-1)+i_2-1)+j_2-1)+i_3\\ m^{n_3}(m(m^{n_2}(m(i_1-1)+j'_1-1)+i_2-1)+j'_2-1)+i_3}}}
\end{align}	
	
One can see that the two lower indices of the elementary E matrices are very similar in the above products. Furthermore, the first lower index takes the values:
	\begin{align*}
	m(&{\red i_1-1})+{\blue j_1-1}+1\\
	m^{n_2}(m(&{\red i_1-1})+{\blue j_1-1})+{\red i_2-1} +1\\
	m(m^{n_2}(m(&{\red i_1-1})+{\blue j_1-1})+{\red i_2-1})+{\blue j_2-1} +1\\
	m^{n_3}(m(m^{n_2}(m(&{\red i_1-1})+{\blue j_1-1})+{\red i_2-1})+{\blue j_2-1})+{\red i_3-1} +1
	\end{align*}	
as the depth of the product is gradually increased.	Sorting the terms by letters, we find for the last computed product that the index takes the values:
	\begin{equation*}
	\begin{array}{rccrcr}
i_3 +	m^{n_3}m m^{n_2}m & (i_1-1) & + &m^{n_3}m m^{n_2}m\cdot m^{-1} & (j_1-1) &  \\
	+ \ m^{n_3}m          & (i_2-1) & + &m^{n_3}m \cdot m^{-1}         & (j_2-1) &   
	\end{array}
	\end{equation*}
Given the above expression, it is convenient to shift the indices of the $A^i$ matrices by one and have them start at 0 rather than 1. Then, the above enabled us to recognize a pattern and led us to conjecture that:
\begin{equation} \label{Eq:Intermediate1}
	 I_m^{\otimes {n_1}} \otimes A^{1} \otimes
	I_m^{\otimes {n_2}} \otimes
	\cdots\otimes A^{Q-1} \otimes
	I_m^{\otimes {n_{Q}}} = \sum_{\bm i \in \Ii} \sum_{\bm j, \bm j' \in \Jj} \bm A_{\bm j,\bm j'} \, E^{(m^L)}_{{\rm ind}(\bm i,\bm j),{\rm ind}(\bm i,\bm j')},
	\end{equation}
where:
\begin{itemize}
\item $\bm A_{\bm j,\bm j'} = A^{1}_{j_1,j'_1} \ldots A^{Q-1}_{j_{Q-1},j'_{Q-1}}$.
\item The multi-index $\bm i = \{i_1,\ldots,i_Q\}$ samples the set:
\begin{equation}\label{Eq:SetI}
\bm i \in \Ii=\{0,\ldots m^{n_1}-1\} \times \ldots \times \{0,\ldots, m^{n_Q}-1\}.
\end{equation}
\item The multi-indices $\bm j=\{j_1,\ldots,j_{Q-1}\}$ and $\bm j' = \{j'_1,\ldots,j'_{Q-1}\}$ sample the set:
\begin{equation}\label{Eq:SetJ}
\bm j,\bm j' \in \Jj=\{0,\ldots,m-1\}^{\times (Q-1)}.
\end{equation}	
\item The indices of the elementary $E$ matrices are supplied by the function $\ind: \Ii \times \Jj \rightarrow \{1,\ldots,m^L\}$: 
	\begin{align}\label{Eq:Ind0}
	\ind(\bm i,\bm j) = i_Q+1+\sum_{r=1}^{Q-1}m^{\left(Q-r+\sum_{s=r+1}^{Q} n_s\right)}(i_r+j_r/m).
	\end{align}
\end{itemize}

The ansatz displayed in \eqref{Eq:Intermediate1} and \eqref{Eq:Ind0} coincides with the expressions derived in \eqref{Eq:T3} and \eqref{Eq:T5} when $Q$ is set at 2 and 3, respectively. Furthermore, we have verified that, after taking the product:
\begin{equation}
\big ( I_m^{\otimes {n_1}} \otimes A^{1} \otimes
	I_m^{\otimes {n_2}} \otimes
	\cdots\otimes A^{Q-1} \otimes
	I_m^{\otimes {n_{Q}}} \big ) \otimes \big (A^{Q} \otimes
	I_m^{\otimes {n_{Q+1}}} \big ),
	\end{equation}
we obtain the same expression \eqref{Eq:Intermediate1} but with $Q$ replaced by $Q+1$. Hence, an induction argument assures us that expression \eqref{Eq:Intermediate1} is valid for all $Q$'s.

\vspace{0.2cm}

\subsection{Final Expressions}
\label{SubSec:FExpr}

For reasons which will soon become apparent, it is convenient to define the function:
\begin{equation}\label{Eq:IndF}
{\rm Ind} : \Ii \times \Jj ^2 \rightarrow \{1,\ldots, m^L\}^{\times 2}, \quad {\rm Ind}(\bm i,\bm j,\bm j') = \big ( \ind(\bm i,\bm j),\ind(\bm i,\bm j')\big )
\end{equation}
and write our expressions as:
	\begin{align} \label{Eq:Intermediate3}
	 I_m^{\otimes {n_1}} \otimes A^{1} \otimes
	I_m^{\otimes {n_2}} \otimes
	\cdots\otimes A^{Q-1} \otimes
	I_m^{\otimes {n_{Q}}} = \sum_{\bm i \in \Ii} \sum_{\bm j, \bm j' \in \Jj} \bm A_{\bm j,\bm j'} \, E^{(m^L)}_{{\rm Ind}(\bm i,\bm j,\bm j')}.
	\end{align}
If $A^i$ are generic $m \times m$ matrices, then the ranges $\Ii$ and $\Jj$  are entirely sampled by the multi-indices. As such, one of the important advantages of \eqref{Eq:Intermediate3} is the precise identification of the indices corresponding to the non-zero entries in the matrix representations of the elementary operators. It is instructive to examine the cardinality of the ranges of these indices. We have:
\begin{equation}
|\Ii| = m^{\sum_{s=1}^Q n_s} = m^{L-Q+1}, \quad |\Jj| = m^{Q-1}.
\end{equation}
As such, the summations in \eqref{Eq:Intermediate3} runs over a total set of cardinality:
\begin{equation} 
|\Ii| \times |\Jj|^2 = m^{L+Q-1}.
\end{equation}
Note that the number of entries in a generic matrix from $\Mm_m^{\otimes L}$ is $m^{2L}$, which in the typical applications is much larger than $m^{L+Q-1}$.

\vspace{0.2cm} 

In many instances, however, the local $A^i$ matrices are sparse and, in such cases, our expressions can be further optimized. Indeed, consider the following sets:
	\begin{equation}
	\Kk_i = \big \{(a,b) \in \{0,\ldots,m - 1 \}^2, \ A^i_{a,b} \neq 0 \big \}, \quad i=1,\ldots,Q-1,
	\end{equation}
	and note that the multi-indices $(\bm j,\bm j')$ in \eqref{Eq:Intermediate3} need to be sampled only over the subset $\Kk_1 \times  \ldots \times \Kk_{Q-1}$ of $\Jj \times \Jj$. Furthermore, note that the sampling of these multi-indices cannot be any smaller than this subset. To take advantage of this important fact, we consider a set of functions:
	\begin{equation}\label{Eq:Gi}
	g_i: \{0,\ldots, |\Kk_i|-1\} \rightarrow \Kk_i, \quad i=1,\ldots,Q-1,
	\end{equation}
	which supply linear labelings for the $(i,j)$ entries of each $\Kk_i$. Then we define the global set:
	\begin{equation}
\Kk=\big \{0,\ldots, |\Kk_1|-1\} \times \ldots \times \{0,\ldots, |\Kk_{Q-1}|-1\big \}
\end{equation}
and the global function:
\begin{equation}
g : \Kk \rightarrow \Jj \times \Jj, \quad g = g_1\times \ldots \times g_{Q-1}.
\end{equation}
Note that the image of the function $g$ is precisely the minimal and optimal sampling set $\Kk_1 \times \ldots \times \Kk_{Q-1}$. Also, note that the ranges of $\bm i$ and $\bm j$ multi-indices in \eqref{Eq:Intermediate3} are independent, hence the order of the two sums can be interchanged. Then we can rewrite \eqref{Eq:Intermediate3} as:
	\begin{align}\label{Eq:Final}
	\boxed{
	 I_m^{\otimes {n_1}} \otimes A^{1} \otimes
	I_m^{\otimes {n_2}} \otimes
	\cdots\otimes A^{Q-1} \otimes
	I_m^{\otimes {n_{Q}}} = \sum_{\bm k \in \Kk} \bm A_{g(\bm k)} \,\sum_{\bm i \in \Ii}  E^{(m^L)}_{{\rm Ind}(\bm i,g(\bm k) )}. }
	\end{align}	
This is our final expression.
	
\vspace{0.2cm}
	
Eq.~\eqref{Eq:Final} takes into account the particularities of matrices $A^i$, which are reflected in the sets $\Kk_i$ and functions $g_i$, and now the summation is over the most restricted set of indices possible. 	Note that the indices in \eqref{Eq:Final} sample a total set of cardinality:
	\begin{equation}
	N = |\Ii | \times |\Kk| = m^{L-Q+1}\times |\Kk_1| \times \ldots \times |\Kk_{Q-1}|.
	\end{equation}
	For example, if $m=2$ and $A^i$ are Pauli matrices, then $N = 2^L$. As such, \eqref{Eq:Final} populates a $2^L \times 2^L$ matrix in just $2^L$ steps.
	
\subsection{Input Data and Soft-Coding}

In the process of our derivation, we in fact systematized the minimal input data required by \eqref{Eq:ElementaryOp}. Indeed, to evaluate \eqref{Eq:ElementaryOp}, we need to specify:
\begin{itemize}
\item The set $\Ii$ or the structure data $Q$ and $\bm n = \{n_1,\ldots,n_{Q-1}\}$.
\item The function $g=g_1 \times \ldots \times g_{Q-1}$, that is, its domain and co-domain, as well as its values.
\item The entries $A^i_{g_i(\bm k)}$ for all $i=1,\ldots,Q-1$ and $\bm k \in \Kk$.
\end{itemize}
Note that once $\bm n$ is known, the function ${\rm Ind}$ defined in \eqref{Eq:IndF} and \eqref{Eq:Ind0} can be evaluated without the need of any additional data. These aspects will be discussed at length in the following section.

\vspace{0.2cm}

The sums in \eqref{Eq:Final} over the multi-indices $\bm k$ and $\bm i$ can be coded as nested loops but such an approach has the disadvantage that these nested loops need to be hard-coded and, as such, the source code will require modifications if, for example, $Q$ is changed. Our solution is to perform the first sum in \eqref{Eq:Final} over a linear global index $\alpha \in \{0,|\Kk|-1\}$:
		\begin{equation}\label{Eq:Alpha}
		\alpha = k_1 + k_2 \, |\Kk_1|+k_3 \, |\Kk_1|\cdot|\Kk_2| + \ldots + k_{Q-1} \, |\Kk_1|\cdot \ldots \cdot |\Kk_{Q-2}|.
\end{equation}
The right hand side is a generalized number base factorization of $\alpha$, hence one can use the following inverse procedure to generate the indices $k_i \in \{0,\ldots,|\Kk_i|-1\}$, hence the multi-index $\bm k = (k_1,\ldots,k_{Q-1}) \in \Kk$, out of the linear index $\alpha$:
\begin{lstlisting}[language=Fortran,caption={ \ },label={Script:1}]
input Q,K(1:Q-1)
input $\alpha \in \{0,|\Kk|-1\}$
allocate kk(1:Q-1)
p=0; w=1; kk(:)=0
do r=1,Q-1
 z=K(r)
 kk(r)=($\alpha$-p)/w mod z 
 p=p+kk(r)*w
 w=w*z
end do
return kk(:)
\end{lstlisting}
Above, the array ${\rm K}(:)$ stores the cardinals $|\Kk_i|$, $i=1,\ldots,Q-1$.
		
\vspace{0.2cm}

As for the second sum in \eqref{Eq:Final}, we will use a linear index $\beta \in \{0,\ldots,|\Ii|-1\}$:
\begin{equation}\label{Eq:Beta}
\beta = i_1 + i_2 \, m^{n_1}+ i_3 \, m^{n_1+n_2} \ldots i_Q \, m^{n_1 + \ldots + n_{Q-1}},
\end{equation} 
and use again a number base conversion to generate the indices $i_r \in (0,m^{n_r}-1)$, hence the mutli-index $\bm i = (i_1,\ldots,i_Q)$. There is one small complication: There are special cases where one or more $n_r$'s could be zero, such as when all $A^i$'s appear at the end of the tensor product. In such cases, the only allowed value of the corresponding indices $i_r$ is zero, which takes them completely out of all equations. As such, we can develop a uniform formalism which treats all such scenarios on equal footing. The script generating the multi-index $\bm i$ is as follows;

\begin{lstlisting}[language=Fortran,caption={ \ },label={Script:2}]
input m,Q,nn(1:Q)
input $\beta \in \{0,{\rm m}^{L-Q+1}-1\}$
allocate ii(1:Q)
p=0; w=1; ii(:)=0
do r=1,Q
 if(nn(r)$\neq$0) then
  z=m**nn(r)
  ii(r)=($\beta$-p)/w mod z 
  p=p+ii(r)*w
  w=w*z
 end if
end do
return ii(:)
\end{lstlisting}
Above, the array ${\rm nn}(:)$ stores the multi-index $\bm n = \{n_1,\ldots,n_Q\}$.

\section{Spin-Hamiltonians over Triangulations}
\label{Sec:SpinH}

This section analyzes spin-Hamiltonians over triangulated surfaces. Specifically, it specifies: 1) how a triangulation and its sub-sets are going to be numerically encoded and how certain structural coefficients are decoded from the input file of a triangulation; 2)  how we structured and coded the input data for a given spin-Hamiltonian; 3) how its matrix elements are numerically evaluated based on \eqref{Eq:Final}. The reader will also find here the main blocks of our source code in a syntax which is close to but not exactly that of Fortran. This was a choice we made so that the reader can see more clearly the connection with the main text. We also believe that this choice will make the translation of the codes in other programming languages easier. At the end of the section, the reader will find an introduction to the complete working Fortran source codes that are supplied in the Appendices. 

\subsection{Triangulations}

Each compact orientable surface $M$ admits a finite triangulation \cite{CarforaBook}, that is, a finite simplicial complex that is homeomorphic to $M$ and such that each edge of the simplex belongs to a 3-leg cycle. Specifically, a triangulation consists of the data $\Ll = (\Vv,\Ee,\Tt)$ specifying its vertices $v \in \Vv$, edges $e \in \Ee$ and elementary triangles $T \in \Tt$. An element of $\Ee$ is an unordered pair $e=\{v,v'\}$ of vertices. Reciprocally, a vertex can be identified with the set $V$ of edges which contain that vertex. Important for our study is ordered labeling of the vertices, edges and elementary triangles, which is assumed automatically. 

\vspace{0.2cm}

The data $\Ll = (\Vv,\Ee,\Tt)$ needs to be accessed, searched and manipulated. Hence, a proper digital encoding of $\Ll$ is very important. The vertices are identified with their labels, hence $\Vv \simeq \big\{1,2,\ldots, |\Vv|\big \}$. An edge $e$ is identified with the subset $e=\{v,v'\} \subset \Vv$ of the supporting vertices. The latter is specified by the function $\eta_e(v)$ returning 1 if $v\in e$ and 0 otherwise, which is numerically implemented by the arrays of $0$'s and $1$'s:
\begin{equation}
{\rm edge}(:,n) = \big (/\eta_n(1),\eta_n(2), \ldots ,\eta_n(|\Vv|) /\big ), \quad n =1,\ldots, |\Ee|.
\end{equation}

\vspace{0.2cm}

It is also important to devise an effective method of encoding and manipulating the subsets $\Gamma$ of $\Ee$. Let us recall that set of all subsets of $\Ee$ is  in one to one correspondence with the set of functions $\chi : \Ee \rightarrow \{0,1\}$, both denoted by the same symbol $2^\Ee$. The explicit correspondence is supplied by:
\begin{equation}
\chi : \Ee \rightarrow \{0,1\} \ \mapsto \ \Gamma=\chi^{-1}(1) \subset \Ee. 
\end{equation}
Numerically, these functions can be specified by the arrays of $0$'s and $1$'s:
\begin{equation}
{\rm chi}(:) = \big (/\chi(1),\chi(2), \ldots ,\chi(|\Ee|) /\big ).
\end{equation} 

\subsection{Unstructured Hamiltonians}	
	
Let $\Ll = (\Vv,\Ee)$ be a triangulation of a surface, where $\Vv$ is the set of vertices and $\Ee$ is the set of edges. To each edge $e$ we associate the local Hilbert spaces $\Hh_e \simeq \CM^m$, hence the global Hilbert space is $\Hh = \bigotimes_{e \in \Ee}  \Hh_e$ and the global algebra of observables is $\bigotimes_{e \in \Ee} \Mm_m$. If $A^e$ is a local operator over $\Hh_e$, we denote by $\hat A^e$ its embedding in the global algebra:
\begin{equation} 
\hat A^e = I_m \otimes \ldots \otimes I_m \otimes A^e \otimes I_m \otimes \ldots \otimes I_m.
\end{equation}
Note that $\hat A^e \hat A^{e'} = \hat A^{e'} \hat A^e$ if $e \neq e'$. Any operator over the global Hilbert space has or it can be brought to the following generic structure:
\begin{equation}\label{Eq:H}
H = \sum_{\Gamma \in 2^\Ee} \sum_{A's}\prod_{e \in \Gamma} \hat A^e.
\end{equation}
As one can see, this results in a sum of elementary operators treated in the previous section.

\vspace{0.2cm}

We assume now that the sets $\Gamma$ were labeled by an index going from 1 to $N_\Gamma$ and that the corresponding functions $\chi_\Gamma$ were all coded as explained in the previous subsection. The structure coefficients $Q(\Gamma)$ and $\bm n(\Gamma)=\big (n_1(\Gamma),\ldots,n_Q(\Gamma)\big )$ entering in \eqref{Eq:Final} and corresponding to each sub-set $\Gamma$ can be determined directly from $\chi_\Gamma$'s as follows:
\begin{lstlisting}[language=Fortran,caption={ \ },label={Script:Structure}]
input $|\Ee|$,N$_\Gamma$,chi(1:$|\Ee|$,1:N$_\Gamma$)\\
allocate(Q(1:N$_\Gamma$))\\
do $\Gamma$=1,N$_\Gamma$\\
 Q($\Gamma$)=sum(chi(:,$\Gamma$))\\
end do
Qmax=max(Q(:))
allocate(nn(1:Qmax,1:N$_\Gamma$))
nn(:,:)=0
do $\Gamma$=1,N$_\Gamma$
 r=1; count=0
 do e=1,$|\Ee|$
  if(chi(e,$\Gamma$)=0) then
   nn(r,$\Gamma$)=nn(r,$\Gamma$)+1
  else
   r=r+1
  end if
 end do
end do
return Q(:),nn(:,:) 
\end{lstlisting}
For a generic Hamiltonian, the maps $g_i$ introduced in \eqref{Eq:Gi} and the non-zero entries $A^i_{a,b}$ need to be hard-coded. In our input files, this part is structured as follows:
\begin{lstlisting}[language=Fortran,caption={ \ },label={Script:6}]
input m,N$_\Gamma$,Q(1:N$_\Gamma$)
allocate(K(1:Qmax-1,1:N$_\Gamma$))
allocate(g1(0:m$^2$-1,1:Qmax-1,1:N$_\Gamma$))
allocate(g2(0:m$^2$-1,1:Qmax-1,1:N$_\Gamma$))
allocate(A(0:m$^2$-1,1:Qmax-1,1:N$_\Gamma$))
K(:,:)=g1(:,:,:)=g2(:,:,:)=A(:,:,:)=0
********************************************
${\rm repeat}$ for $\Gamma$=1,...,N$_\Gamma$ ${\rm and}$ r=1,...,Q($\Gamma$)-1
********************************************
K(r,$\Gamma$) = $|\Kk_r(\Gamma)|$
g1(0:K(r,$\Gamma$)-1,r,$\Gamma$) = (/$\alpha^r_0,\ldots,\alpha^r_{{\rm K(r,\Gamma)}-1}$/)
g2(0:K(r,$\Gamma$)-1,r,$\Gamma$) = (/$\beta^r_0,\ldots,\beta^r_{{\rm K(r,\Gamma)}-1}$/)
A(0:K(r,$\Gamma$)-1,r,$\Gamma$) = (/$A^r_{\alpha^r_0,\beta^r_0}(\Gamma),\ldots,A^r_{\alpha^r_{{\rm K(r,\Gamma)}-1},\beta^r_{{\rm K(r,\Gamma)}-1}}(\Gamma)$/)
*********************************************
\end{lstlisting}
The $\alpha$'s, $\beta$'s and $A$'s at the right side of the last three lines are hard numbers derived from a particular Hamiltonian \eqref{Eq:H}. Also, the arrays g1(:,r,$\Gamma$) and g2(:,r,$\Gamma$) store the two indices $a$ and $b$ outputted by the function $g_r$ defined in \eqref{Eq:Gi}. Note that we made the choice to keep the size of the arrays the same, even though they can be tightly tailored for each $\Gamma$. This is because the sizes are small and an optimization will make no difference while it will complicate the code.

\subsection{Uniformly Structured Hamiltonians}	

For Kitaev's toric model treated in the next section, the subsets $\Gamma$ are restricted to the vertices $V$ and triangles $T$ of a triangulation. Furthermore, the local $A^e$ operators coincide with Pauli matrix $\sigma_3$ for all $V \in \Vv$ and with Pauli matrix $\sigma_1$ for all $T \in \Tt$. In this case, $|\Kk_i(\Gamma)|$ are all equal to 2 and the functions $g_i$ can be soft-coded. Hence, for such applications, the input lines of our code are structured as it follows:

\begin{lstlisting}[language=Fortran,caption={ \ },label={Script:StructuredH}]
input m,N$_\Gamma$,Q(1:N$_\Gamma$)
allocate(K(1:Qmax-1,1:N$_\Gamma$))
allocate(g1(0:m$^2$-1,1:Qmax-1,1:N$_\Gamma$))
allocate(g2(0:m$^2$-1,1:Qmax-1,1:N$_\Gamma$))
allocate(A(0:m$^2$-1,1:Qmax-1,1:N$_\Gamma$))
K(:,:)=g1(:,:,:)=g2(:,:,:)=A(:,:,:)=0
do $\Gamma$=1,N$_\Gamma$
 do r=1,Q($\Gamma$)-1
  K(r,$\Gamma$)=2
  if ($\Gamma \leq |\Vv|$) then   
   g1(0,r,$\Gamma$)=1; g2(0,r,$\Gamma$)=1; A(0,r,$\Gamma$)=1
   g1(1,r,$\Gamma$)=2; g2(1,r,$\Gamma$)=2; A(1,r,$\Gamma$)=-1
  else  
   g1(0,r,$\Gamma$)=1; g2(0,r,$\Gamma$)=2; A(0,r,$\Gamma$)=1
   g1(1,r,$\Gamma$)=2; g2(1,r,$\Gamma$)=1; A(1,r,$\Gamma$)=1
  end if
 end do
end do
\end{lstlisting}
Note that, above, it is assumed that the sets $\Gamma$ corresponding to the vertices are placed at the beginning of the input file. This assumption will be enforced for all our input files.

\subsection{Mixedly Structured Hamiltonians}

The most common situation found in applications is that of mixedly structured Hamiltonians. Indeed, for Kitaev toric model, the main interest is in the braiding of topological defects. These topological defects appear as additional unstructured terms in the Hamiltonian. As such, we adapt our source code to a situation where most of the input data is structured but there is a small number of subsets $\Gamma$ for which the input data needs to be hard coded. 

\vspace{0.2cm}

For the Kitaev toric model with defects, the input lines will be structured as it follows:

\begin{lstlisting}[language=Fortran,caption={\ },label={Script:UnStructuredH}]
input m,N$_\Gamma$,Q(1:N$_\Gamma$)
allocate(K(1:Qmax-1,1:N$_\Gamma$))
allocate(g1(0:m$^2$-1,1:Qmax-1,1:N$_\Gamma$))
allocate(g2(0:m$^2$-1,1:Qmax-1,1:N$_\Gamma$))
allocate(A(0:m$^2$-1,1:Qmax-1,1:N$_\Gamma$))
K(:,:)=g1(:,:,:)=g2(:,:,:)=A(:,:,:)=0
do $\Gamma$=1,$|\Vv|$+$|\Tt|$
 do r=1,Q($\Gamma$)-1
  K(r,$\Gamma$)=2  
  if ($\Gamma \leq |\Vv|$) then
   g1(0,r,$\Gamma$)=1; g2(0,r,$\Gamma$)=1; A(0,r,$\Gamma$)=1
   g1(1,r,$\Gamma$)=2; g2(1,r,$\Gamma$)=2; A(1,r,$\Gamma$)=-1
  else  
   g1(0,r,$\Gamma$)=1; g2(0,r,$\Gamma$)=2; A(0,r,$\Gamma$)=1
   g1(1,r,$\Gamma$)=2; g2(1,r,$\Gamma$)=1; A(1,r,$\Gamma$)=1
  end if
 end do
end do
**********************************************
${\rm repeat}$ for $\Gamma$=$|\Vv|$+$|\Tt|$+1,...,N$_\Gamma$ ${\rm and}$ r=1,...,Q($\Gamma$)-1
**********************************************
K(r,$\Gamma$) = $|\Kk_r(\Gamma)|$
g1(0:K(r,$\Gamma$)-1,r,$\Gamma$) = (/$\alpha^r_0,\ldots,\alpha^r_{{\rm K(r,\Gamma)}-1}$/)
g2(0:K(r,$\Gamma$)-1,r,$\Gamma$) = (/$\beta^r_0,\ldots,\beta^r_{{\rm K(r,\Gamma)}-1}$/)
A(0:K(r,$\Gamma$)-1,r,$\Gamma$) = (/$A^r_{\alpha^r_0,\beta^r_0}(\Gamma),\ldots,A^r_{\alpha^r_{{\rm K(r,\Gamma)}-1},\beta^r_{{\rm K(r,\Gamma)}-1}}(\Gamma)$/)
**********************************************
\end{lstlisting}
Testing and simulation of such cases are left to the future.

\subsection{Generating the Matrix Elements}

The matrix elements of the elementary terms of the Hamiltonian are written explicitly in \eqref{Eq:Final}. We reproduce them below, this time emphasizing the dependence on the subsets $\Gamma$:
	\begin{align}\label{Eq:Ele}
	& I_m^{\otimes {n_1(\Gamma)}} \otimes A^{1}(\Gamma) \otimes
	I_m^{\otimes {n_2}(\Gamma)} \otimes
	\cdots\otimes A^{Q(\Gamma)-1}(\Gamma) \otimes
	I_m^{\otimes {n_{Q(\Gamma)}(\Gamma)}} \\ \nonumber 
	& \qquad = \sum_{\bm k \in \Kk(\Gamma)} \bm A_{g(\bm k;\Gamma)}(\Gamma) \,\sum_{\bm i \in \Ii}  E^{(m^L)}_{{\rm Ind}(\bm i,g(\bm k;\Gamma) )}. 
	\end{align}	
The function $g=g_1 \times \ldots \times g_{Q(\Gamma)-1}$ was already digitized in the previous section and the function ${\rm Ind}$ is defined in \eqref{Eq:IndF}. Furthermore:
\begin{equation}
\bm A_{g(\bm k;\Gamma)}(\Gamma)=\prod_{r=1}^{Q(\Gamma)-1} A^r_{g_r(k_r,\Gamma)}(\Gamma)
\end{equation}
Let us also recall that the multi-indices $\bm k$ and $\bm i$ are encoded in the linear indices $\alpha$ and $\beta$ defined in Eqs.~\ref{Eq:Alpha} and \ref{Eq:Beta}, respectively. The expression \eqref{Eq:Ele} was implemented in our code as it follows.

\begin{lstlisting}[language=Fortran,caption={ \ },label={Script:9}]
input m,$|\Ee|$,chi(1:$|\Ee|$,1:N$_\Gamma$)
insert Block $\ref{Script:Structure}$
insert Block $\ref{Script:UnStructuredH}$
Ham=0
do $\Gamma$=1,Ng
 do $\alpha$=0,Nk($\Gamma$)-1
  insert Block $\ref{Script:1}$
  ${\bm A}$=1d0
  do r=1,Q($\Gamma$)-1
   ${\bm A}$=${\bm A}$*A(kk(r),r,$\Gamma$)
  end do
  do $\beta$=0,m**($|\Ee|$-Q($\Gamma$)+1)-1
   insert Block $\ref{Script:2}$
   Ind1=Ind1+ii(Q($\Gamma$))+1
   Ind2=Ind2+ii(Q($\Gamma$))+1
   do r=1,Q(ga)-1
    expo=sum(nn(r+1:Q($\Gamma$),$\Gamma$))+Q($\Gamma$)-r
    Ind1=Ind1+(ii(r)+g1(kk(r),r,$\Gamma$)/m)*m**expo
    Ind2=Ind2+(ii(r)+g2(kk(r),r,$\Gamma$)/m)*m**expo
   end do
   Ham(Ind1,Ind2)=Ham(Ind1,Ind2)+${\bm A}$
  end do
 end do
end do
\end{lstlisting}

\vspace{0.2cm}

A complete Fortran script that computes and diagonalizes the full Hamiltonian \eqref{Eq:KHam} is supplied in \ref{Script:FullCode1}. The code is specialized for the un-perturbed Kitaev's model \eqref{Eq:KHam}, which is investigated in the next section. As such, the script contains the block \ref{Script:StructuredH}, which sets the functions $g$ to values specific to this particular model. A scientist interested in other spin models will have to re-code this section of the script. Also, if the scientist wants to investigate topological defects, then block~\ref{Script:StructuredH} needs to be promoted to block~\ref{Script:UnStructuredH}. The input files are supplied in the next section for various triangulations of interest.

\vspace{0.2cm}

Even for modestly sized triangulations, the Hilbert space dimension, $2^{|\Ee|}$, is extremely large and the full diagonalization of \eqref{Eq:KHam} becomes un-feasible. Furthermore, the main interest in most applications is in the low energy spectral properties of the Kitaev model. As such, we supply in \ref{Script:FullCode2} and \ref{Script:FullCode3} two Fortran scripts that resolves the lowest $p$ eigenvalues and their corresponding eigenvectors. The algorithm, which is based on minimizing ${\rm Tr}(\Pi_p H \Pi_p)$ with respect to the rank-$p$ projection $\Pi_p$, was adapted from and described in previous large-scale electronic structure projects \cite{ProdanNL2008}. Let us point out that \ref{Script:FullCode2} and \ref{Script:FullCode3} require only the action of the Hamiltonian on vectors, which is computed on the fly and no storage of $H$ itself is needed. As already argued in section~\ref{SubSec:FExpr}, our close-form expressions supply the optimal coding of this Hamiltonian action on vectors. The algorithm in \ref{Script:FullCode2} proceeds iteratively by determining the best candidate $\Pi_p^{(j+1)}$ of $\Pi_p$ inside the $2p$-dimensional space $\Pi_p^{(j)}\Hh + H\Pi_p^{(j)}\Hh$, where $\Hh$ is the global Hilbert space. $\Pi_p^{(0)}$ is initiated with random entries and the iteration exits when the spaces $\Pi_p^{(j)}\Hh$  and $H\Pi_p^{(j)}\Hh$ become identical up to an $\epsilon$ tolerance. We call $\epsilon$ the stop parameter.

\vspace{0.2cm}

The difference between the two source codes \ref{Script:FullCode2} and \ref{Script:FullCode3} is in how the parallelization was implemented. Both scripts contain simple OpenMP directives that parallelize the large loops over $\Gamma$. Inside these loops, information is loaded, processed and dumped on the same memory location. To avoid any interference between the parallel workloads, in \ref{Script:FullCode2}, the code was modified such that the loops over $\Gamma$ become fully independent at the expense of memory consumption.  In \ref{Script:FullCode3}, this safety feature has been removed for the second large loop over $\Gamma$ and, as a result, the memory consumption is highly reduced for this version of the code. The two source codes were extensively tested and found to virtually return identical outputs.

\section{Applications}
\label{Sec:Applications}

Our application consists of a numerical demonstration of the topological degeneracy for Kitaev's toric model \cite{KitaevAOP2003,KitaevAOP2006}. We singled out this model and this application for the following reasons: 
\begin{itemize}
\item The model played and continues to play a central role in the field of topological quantum computation \cite{KitaevOxford2008}. As such, the model is of interest to a large community of scientists yet its numerical investigations are scarce; 
\item 
The Hamiltonians related to the model are complex as they involve higher rank tensor products. As such, they present the kind of challenges our code was designed for; 
\item There are precise analytic predictions about the model, which can be used to test our codes.
\item While many statements about the unperturbed toric model can be obtained analytically, its properties under strong perturbations are largely unknown and sorting them out will require numerical simulations. Currently, there is a strong interest in this aspect and interesting predictions have been already put forward in the published literature \cite{BrunoArxiv2019}.
\end{itemize}

\vspace{0.2cm}

The model is defined as follows. Let $\Ll=(\Vv,\Ee)$ be a triangulation of a closed orientable surface, where $\Vv$ is the set of vertices and $\Ee$ is the set of edges. A vertex will be identified with the set of edges emanating from it. To each edge $e$, one attaches  a Hilbert space $\Hh_e \simeq \CM^2$ and then forms the global Hilbert space of the model $\Hh = \bigotimes_{e \in \Ee} \Hh_e$, where an ordering is assumed on $\Ee$. On this Hilbert space, there are the operators:
\begin{equation}
X_e = I_2 \otimes \ldots \otimes I_2 \otimes \sigma_x \otimes I_2 \ldots \otimes I_2
\end{equation}
and
\begin{equation}
Z_e = I_2 \otimes \ldots \otimes I_2 \otimes \sigma_z \otimes I_2 \ldots \otimes I_2,
\end{equation}
where $\sigma$'s are Pauli's matrices. Above, the Pauli matrices $\sigma_x$ and $\sigma_z$ sit on edge $e$. With these operators, one forms:
\begin{equation}
A_V = \prod_{e \in V} X_e, \quad B_T = \prod_{e \in T} Z_e, \quad V\in \Vv, \quad T\in \Tt.
\end{equation}
Lastly, the Hamiltonian of the model is:
\begin{equation}\label{Eq:KHam}
H = - \sum_{V \in \Ll} A_V - \sum_{T \in \Tt} B_T.
\end{equation} 

\vspace{0.2cm}

Among the theoretical predictions on the above Hamiltonian, one will find the following statements:
\begin{itemize}
\item The spectrum of $H$ consists of energy levels equally spaced by 4 units.
\item The ground state energy is at $E_0 = -(|\Vv| + |\Tt|)$.
\item The degeneracy of the ground state is $4^g$, where $g$ is the genus of the triangulated surface.
\end{itemize}

\vspace{0.2cm}

\begin{figure}[t]
    	\centering
    	\includegraphics[width=0.75\linewidth]{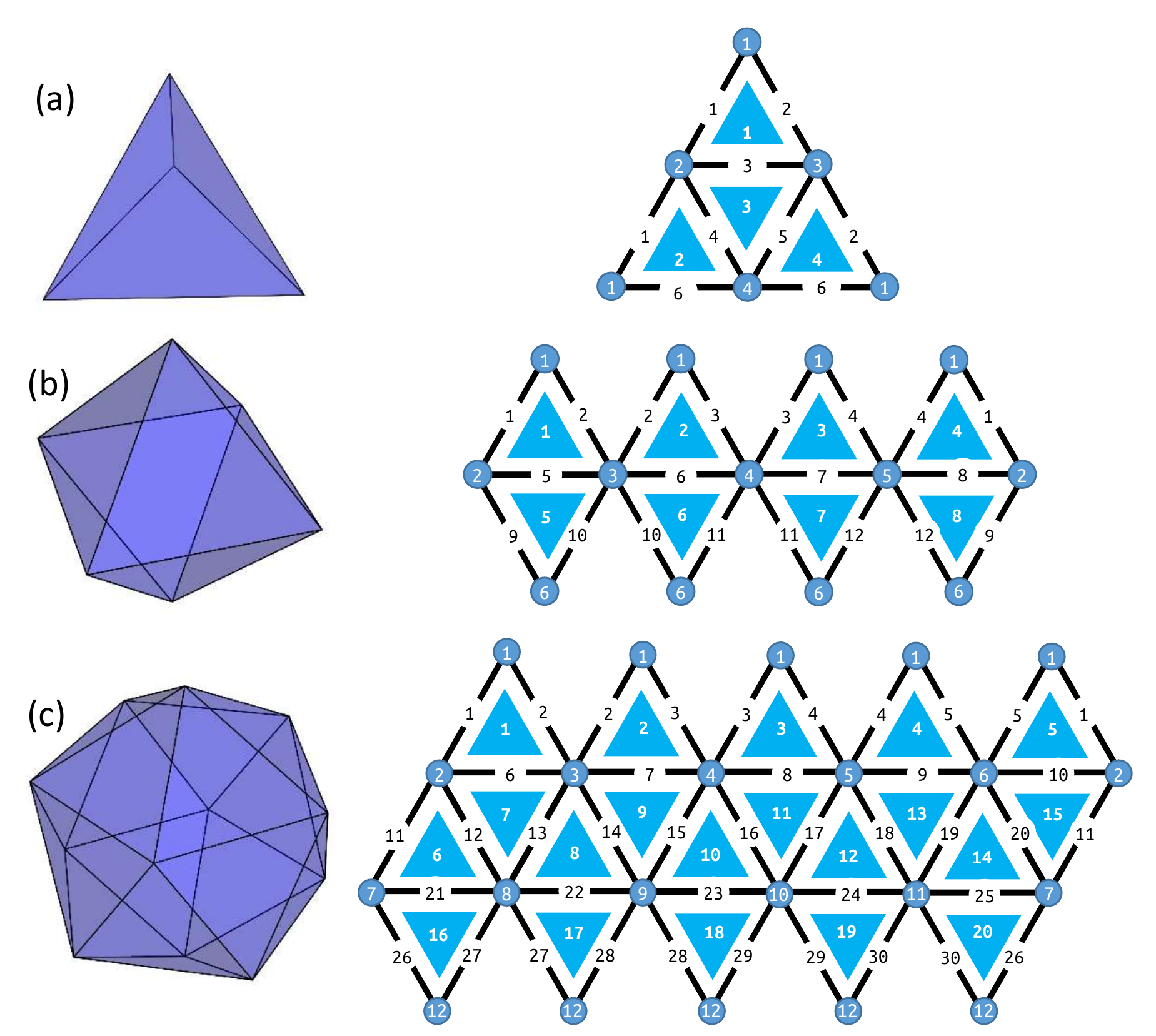}
    	\caption{Triangulations of the sphere by regular polyhedra and their corresponding labelings of the vertices, edges and triangles: (a) Tetrahedron triangulation; (b) Octahedron triangulation; (c) Icosahedron triangulation.} 
    	\label{Fig:Sphere}
    \end{figure}

\begin{figure}[t]
\centering
\includegraphics[width=0.72\linewidth]{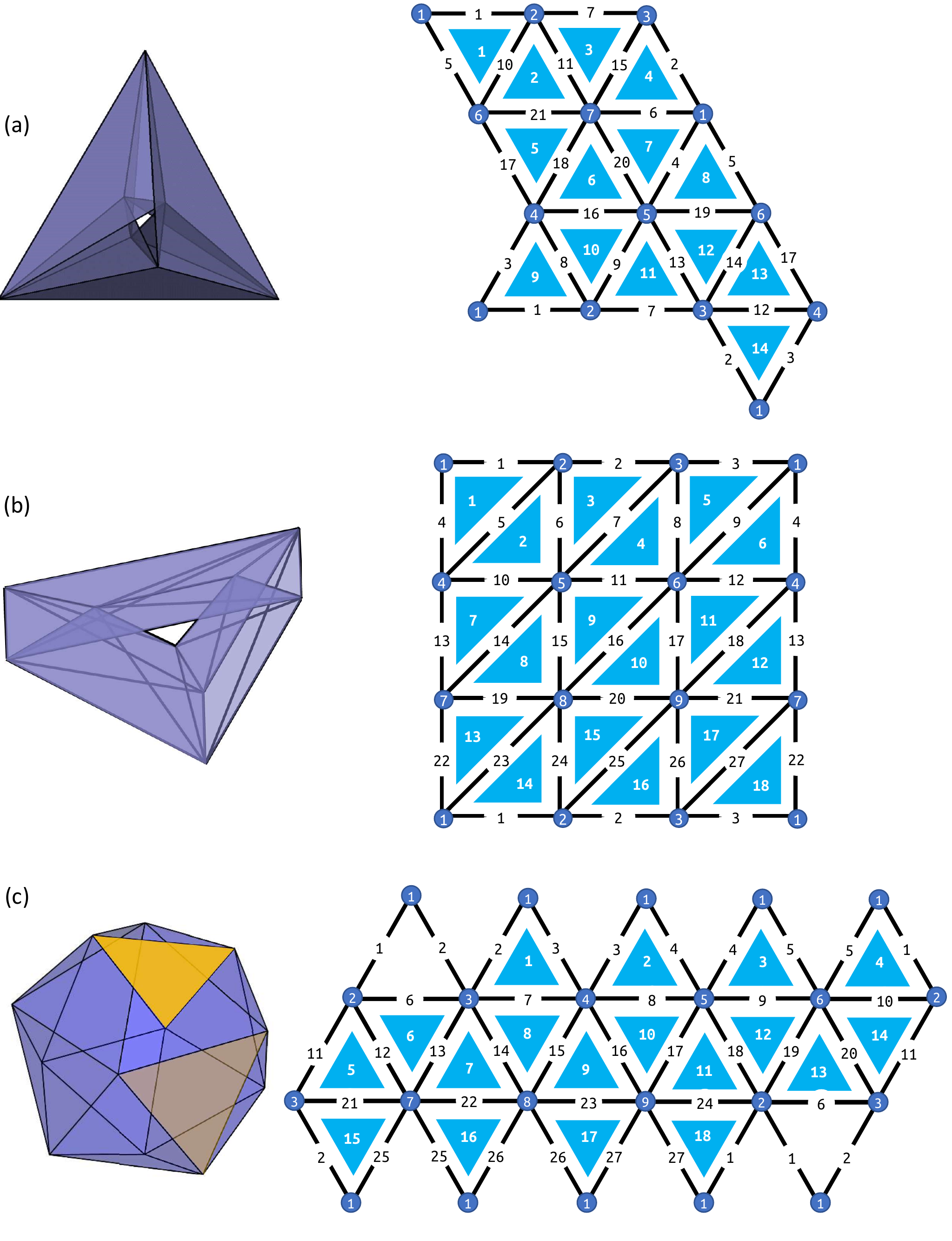}
\caption{Triangulations of the 1-torus and their corresponding labelings of the vertices, edges and triangles: (a) Minimal triangulation; (b) Translation invariant triangulation; (c) Triangulation generated by pinching the icosahedron at the two shaded triangles.} 
\label{Fig:Torus}
\end{figure}

These are the statements that we will numerically confirm in this section for the set of triangulations shown in Figs.~\ref{Fig:Sphere} and \ref{Fig:Torus}. Note the complexity of the vertex operators for these triangulations, which consist of products of 6 spin operators. Let us point out the following useful information from \cite{JungermanAM1980} about the minimal triangulation of a closed orientable surface of genus $g$, which says that the number of triangles $N_t$ in a minimal triangulation is given by:
\begin{equation}
N_t^{\rm min} = 2\Big \{\tfrac{7+\sqrt{1+48 g}}{2} \Big \}+4(g-1), \quad g \neq 2,
\end{equation}
and $N_t^{\rm min}=24$ if $g=2$. The symbol $\{x\}$ denotes the smallest integer greater or equal to $x$. According to this formula, the minimal triangulations will contain 4, 14 and 24 triangles for $g=0$, 1 and 2, respectively. One will also find that the numbers of edges for these minimal triangulations are  6, 21 and 36 for $g=0$, 1 and 2, respectively. This means that the minimal Kitaev model on a 2-torus is defined on a Hilbert space of dimension $2^{36}$. Beyond $g=2$, the dimensions of the Hilbert spaces corresponding to the minimal triangulations are staggering and this highlights again the numerical challenges related to the toric model. 

\subsection{Orientable triangulations of genus 0}

The orientable surfaces of genus 0 are all topologically equivalent to the sphere. The minimal triangulation of the sphere contains 4 triangles and it is supply by the tetrahedron \cite{JungermanAM1980}, as shown in Fig.~\ref{Fig:Sphere}(a). The input file for this triangulation is supplied below:

\begin{lstlisting}[caption={\ },label={Script:Thetrahedron}]
Nv=4;Ne=6;Nt=4;Ng=Nv+Nt 
allocate(chi(1:Ne,1:Ng))
!!!!vertex data!!!! 
chi(:,1)=(/1,1,0,0,1,0/)
chi(:,2)=(/1,0,1,1,0,0/)
chi(:,3)=(/0,1,1,0,0,1/)
chi(:,4)=(/0,0,0,1,1,1/)
!!!!triangle data!!!! 
chi(:,5)=(/1,1,1,0,0,0/)
chi(:,6)=(/1,0,0,1,1,0/)
chi(:,7)=(/0,1,0,0,1,1/)
chi(:,8)=(/0,0,1,1,0,1/)
\end{lstlisting}
The Hilbert space dimension is $2^6$, small enough to permit a computation and diagonalization of the full Hamiltonian, even on a laptop. With the above input, the code supplied in \ref{Script:FullCode1} returns a singly degenerate eigenvalue at $-8$, a 12-fold degenerate eigenvalue at -4, a 38-fold degenerate eigenvalue at 0, a 12-fold degenerate eigenvalue at 4 and a non-degenerate eigenvalue at 8. The values of these eigenvalues are exact to 15 digits of precision. As a test, we also ran the code supplied in \ref{Script:FullCode2}, which resolves only the lower part of the spectrum. We found that, with the stop parameter $\epsilon = 10^{-4}$, \ref{Script:FullCode2} reproduces the exact spectrum with 4 digits of precision, for the first three energy levels. The degeneracy of these energy levels is also correctly reproduced.

\vspace{0.2cm}

The triangulation supplied by the octahedron is shown in Fig.~\ref{Fig:Sphere}(b) and the corresponding input file is supplied below:
\begin{lstlisting}[caption={\ },label={Script:Octahedron}]
Nv=6;Ne=12;Nt=8;Ng=Nv+Nt
allocate(chi(1:Ne,1:Ng))
!!!!vertex data!!!!
 chi(:,1)=(/1,1,1,1,0,0,0,0,0,0,0,0/)
 chi(:,2)=(/1,0,0,0,1,0,0,1,1,0,0,0/)
 chi(:,3)=(/0,1,0,0,1,1,0,0,0,1,0,0/)
 chi(:,4)=(/0,0,1,0,0,1,1,0,0,0,1,0/)
 chi(:,5)=(/0,0,0,1,0,0,1,1,0,0,0,1/)
 chi(:,6)=(/0,0,0,0,0,0,0,0,1,1,1,1/)
!!!!triangle data!!!!
 chi(:,7)=(/1,1,0,0,1,0,0,0,0,0,0,0/)
 chi(:,8)=(/0,1,1,0,0,1,0,0,0,0,0,0/)
 chi(:,9)=(/0,0,1,1,0,0,1,0,0,0,0,0/)
chi(:,10)=(/1,0,0,1,0,0,0,1,0,0,0,0/)
chi(:,11)=(/0,0,0,0,1,0,0,0,1,1,0,0/)
chi(:,12)=(/0,0,0,0,0,1,0,0,0,1,1,0/)
chi(:,13)=(/0,0,0,0,0,0,1,0,0,0,1,1/)
chi(:,14)=(/0,0,0,0,0,0,0,1,1,0,0,1/)
\end{lstlisting}
The Hilbert space dimension is $2^{12}$, still small enough to permit a computation and diagonalization of the full Hamiltonian and such task can be still  carried on a laptop. With the above input, the code supplied in \ref{Script:FullCode1} returns a singly degenerate eigenvalue at $-14$, a 43-fold degenerate eigenvalue at -10, a 505-fold degenerate eigenvalue at -6, a 1499-fold degenerate eigenvalue at -2, a 1499-fold degenerate eigenvalue at 2, 505 -fold degenerate eigenvalue at 6, a 43-fold degenerate eigenvalue at 10 and a non-degenerate eigenvalue at 14. The values of these eigenvalues are exact to 15 digits of precision. As a test, we also ran the source codes supplied in \ref{Script:FullCode2} and \ref{Script:FullCode3}. With the stop parameter $\epsilon = 10^{-4}$, they reproduced the exact ground state energy and the value of the spectral gap with 4 digits of precision, as well as the non-degenerate character of the ground state.

\vspace{0.2cm}

The triangulation supplied by the icosahedron is shown in Fig.~\ref{Fig:Sphere}(c) and the corresponding input file is supplied below:

\begin{lstlisting}[caption={\ },label={Script:Icosahedron}]
Nv=12;Ne=30;Nt=20;Ng=Nv+Nt
allocate(chi(1:Ne,1:Ng))
!!!!vertex data!!!!
 chi(:,1)=(/1,1,1,1,1,0,0,0,0,0,0,0,0,0,0,0,0,0,0,0,0,0,0,0,0,0,0,0,0,0/)
 chi(:,2)=(/1,0,0,0,0,1,0,0,0,1,1,1,0,0,0,0,0,0,0,0,0,0,0,0,0,0,0,0,0,0/)
 chi(:,3)=(/0,1,0,0,0,1,1,0,0,0,0,0,1,1,0,0,0,0,0,0,0,0,0,0,0,0,0,0,0,0/)
 chi(:,4)=(/0,0,1,0,0,0,1,1,0,0,0,0,0,0,1,1,0,0,0,0,0,0,0,0,0,0,0,0,0,0/)
 chi(:,5)=(/0,0,0,1,0,0,0,1,1,0,0,0,0,0,0,0,1,1,0,0,0,0,0,0,0,0,0,0,0,0/)
 chi(:,6)=(/0,0,0,0,1,0,0,0,1,1,0,0,0,0,0,0,0,0,1,1,0,0,0,0,0,0,0,0,0,0/)
 chi(:,7)=(/0,0,0,0,0,0,0,0,0,0,1,0,0,0,0,0,0,0,0,1,1,0,0,0,1,1,0,0,0,0/)
 chi(:,8)=(/0,0,0,0,0,0,0,0,0,0,0,1,1,0,0,0,0,0,0,0,1,1,0,0,0,0,1,0,0,0/)
 chi(:,9)=(/0,0,0,0,0,0,0,0,0,0,0,0,0,1,1,0,0,0,0,0,0,1,1,0,0,0,0,1,0,0/)
chi(:,10)=(/0,0,0,0,0,0,0,0,0,0,0,0,0,0,0,1,1,0,0,0,0,0,1,1,0,0,0,0,1,0/)
chi(:,11)=(/0,0,0,0,0,0,0,0,0,0,0,0,0,0,0,0,0,1,1,0,0,0,0,1,1,0,0,0,0,1/)
chi(:,12)=(/0,0,0,0,0,0,0,0,0,0,0,0,0,0,0,0,0,0,0,0,0,0,0,0,0,1,1,1,1,1/)
!!!!triangle data!!!!
chi(:,13)=(/1,1,0,0,0,1,0,0,0,0,0,0,0,0,0,0,0,0,0,0,0,0,0,0,0,0,0,0,0,0/)
chi(:,14)=(/0,1,1,0,0,0,1,0,0,0,0,0,0,0,0,0,0,0,0,0,0,0,0,0,0,0,0,0,0,0/)
chi(:,15)=(/0,0,1,1,0,0,0,1,0,0,0,0,0,0,0,0,0,0,0,0,0,0,0,0,0,0,0,0,0,0/)
chi(:,16)=(/0,0,0,1,1,0,0,0,1,0,0,0,0,0,0,0,0,0,0,0,0,0,0,0,0,0,0,0,0,0/)
chi(:,17)=(/1,0,0,0,1,0,0,0,0,1,0,0,0,0,0,0,0,0,0,0,0,0,0,0,0,0,0,0,0,0/)
chi(:,18)=(/0,0,0,0,0,0,0,0,0,0,1,1,0,0,0,0,0,0,0,0,1,0,0,0,0,0,0,0,0,0/)
chi(:,19)=(/0,0,0,0,0,1,0,0,0,0,0,1,1,0,0,0,0,0,0,0,0,0,0,0,0,0,0,0,0,0/)
chi(:,20)=(/0,0,0,0,0,0,0,0,0,0,0,0,1,1,0,0,0,0,0,0,0,1,0,0,0,0,0,0,0,0/)
chi(:,21)=(/0,0,0,0,0,0,1,0,0,0,0,0,0,1,1,0,0,0,0,0,0,0,0,0,0,0,0,0,0,0/)
chi(:,22)=(/0,0,0,0,0,0,0,0,0,0,0,0,0,0,1,1,0,0,0,0,0,0,1,0,0,0,0,0,0,0/)
chi(:,23)=(/0,0,0,0,0,0,0,1,0,0,0,0,0,0,0,1,1,0,0,0,0,0,0,0,0,0,0,0,0,0/)
chi(:,24)=(/0,0,0,0,0,0,0,0,0,0,0,0,0,0,0,0,1,1,0,0,0,0,0,1,0,0,0,0,0,0/)
chi(:,25)=(/0,0,0,0,0,0,0,0,1,0,0,0,0,0,0,0,0,1,1,0,0,0,0,0,0,0,0,0,0,0/)
chi(:,26)=(/0,0,0,0,0,0,0,0,0,0,0,0,0,0,0,0,0,0,1,1,0,0,0,0,1,0,0,0,0,0/)
chi(:,27)=(/0,0,0,0,0,0,0,0,0,1,1,0,0,0,0,0,0,0,0,1,0,0,0,0,0,0,0,0,0,0/)
chi(:,28)=(/0,0,0,0,0,0,0,0,0,0,0,0,0,0,0,0,0,0,0,0,1,0,0,0,0,1,1,0,0,0/)
chi(:,29)=(/0,0,0,0,0,0,0,0,0,0,0,0,0,0,0,0,0,0,0,0,0,1,0,0,0,0,1,1,0,0/)
chi(:,30)=(/0,0,0,0,0,0,0,0,0,0,0,0,0,0,0,0,0,0,0,0,0,0,1,0,0,0,0,1,1,0/)
chi(:,31)=(/0,0,0,0,0,0,0,0,0,0,0,0,0,0,0,0,0,0,0,0,0,0,0,1,0,0,0,0,1,1/)
chi(:,32)=(/0,0,0,0,0,0,0,0,0,0,0,0,0,0,0,0,0,0,0,0,0,0,0,0,1,1,0,0,0,1/)
\end{lstlisting}

In this case, the Hilbert space dimension is $2^{30}$ and computing and diagonalizing the entire Hamiltonian will require a supercomputer. However, using the code supplied in \ref{Script:FullCode2}, we were able to resolve the ground state energy and the corresponding eigenvector using a workstation with 128 CPUs and 512Gb of shared memory. With the stop parameter set at $\epsilon = 10^{-4}$, the iteration exited at step 36 and returned the ground state energy -31.9872722646278. With the source code supplied in \ref{Script:FullCode3}, we were able to also resolve  the first excited level energy and one associated eigenvector, using the same workstation. With the stop parameter set at $\epsilon=10^{-4}$, the iteration exited at step 38 and returned the eigenvalues -31.9934611540635 and -27.9901315963859.

\subsection{Orientable triangulations of genus 1}

The minimal triangulation of the 1-torus is shown in Fig.~\ref{Fig:Torus}(a) and corresponding input file is supplied below:

\begin{lstlisting}[caption={\ },label={Script:ATorus}]
Nv=7;Ne=21;Nt=14;Ng=Nv+Nt
allocate(chi(1:Ne,1:Ng))
!!!!vertex data!!!!
 chi(:,1)=(/1,1,1,1,1,1,0,0,0,0,0,0,0,0,0,0,0,0,0,0,0/)
 chi(:,2)=(/1,0,0,0,0,0,1,1,1,1,1,0,0,0,0,0,0,0,0,0,0/)
 chi(:,3)=(/0,1,0,0,0,0,1,0,0,0,0,1,1,1,1,0,0,0,0,0,0/)
 chi(:,4)=(/0,0,1,0,0,0,0,1,0,0,0,1,0,0,0,1,1,1,0,0,0/)
 chi(:,5)=(/0,0,0,1,0,0,0,0,1,0,0,0,1,0,0,1,0,0,1,1,0/)
 chi(:,6)=(/0,0,0,0,1,0,0,0,0,1,0,0,0,1,0,0,1,0,1,0,1/)
 chi(:,7)=(/0,0,0,0,0,1,0,0,0,0,1,0,0,0,1,0,0,1,0,1,1/)
!!!!triangle data!!!!
 chi(:,8)=(/1,0,0,0,1,0,0,0,0,1,0,0,0,0,0,0,0,0,0,0,0/)
 chi(:,9)=(/0,0,0,0,0,0,0,0,0,1,1,0,0,0,0,0,0,0,0,0,1/)
chi(:,10)=(/0,0,0,0,0,0,1,0,0,0,1,0,0,0,1,0,0,0,0,0,0/)
chi(:,11)=(/0,1,0,0,0,1,0,0,0,0,0,0,0,0,1,0,0,0,0,0,0/)
chi(:,12)=(/0,0,0,0,0,0,0,0,0,0,0,0,0,0,0,0,1,1,0,0,1/)
chi(:,13)=(/0,0,0,0,0,0,0,0,0,0,0,0,0,0,0,1,0,1,0,1,0/)
chi(:,14)=(/0,0,0,1,0,1,0,0,0,0,0,0,0,0,0,0,0,0,0,1,0/)
chi(:,15)=(/0,0,0,1,1,0,0,0,0,0,0,0,0,0,0,0,0,0,1,0,0/)
chi(:,16)=(/1,0,1,0,0,0,0,1,0,0,0,0,0,0,0,0,0,0,0,0,0/)
chi(:,17)=(/0,0,0,0,0,0,0,1,1,0,0,0,0,0,0,1,0,0,0,0,0/)
chi(:,18)=(/0,0,0,0,0,0,1,0,1,0,0,0,1,0,0,0,0,0,0,0,0/)
chi(:,19)=(/0,0,0,0,0,0,0,0,0,0,0,0,1,1,0,0,0,0,1,0,0/)
chi(:,20)=(/0,0,0,0,0,0,0,0,0,0,0,1,0,1,0,0,1,0,0,0,0/)
chi(:,21)=(/0,1,1,0,0,0,0,0,0,0,0,1,0,0,0,0,0,0,0,0,0/)
\end{lstlisting}
The Hilbert space dimension is $2^{21}$, too large to permit a computation and diagonalization of the full Hamiltonian without substantial computational resources. However, the first five eigenvalues and corresponding eigenvectors can be resolved using the code supplied in \ref{Script:FullCode2}. With the stop parameter $\epsilon = 10^{-4}$, the iteration exited at the 39st step and returned the following eigenvalues: -20.9999999986959,       -20.9999065640024, -20.9997895551610, -20.9957261123087 and       -16.9999990576055, which is in full agreement with the theoretical predictions. In particular, the calculation confirms the expected topological degeneracy. With same input, the source code \ref{Script:FullCode3} was executed on a laptop and the output was virtually identical.

\vspace{0.2cm}

A more regular but larger triangulation of the 1-torus is shown in Fig.~\ref{Fig:Torus}(b) and its corresponding input file is supplied below.
\begin{lstlisting}[caption={\ },label={Script:BTorus}]
Nv=9;Ne=27;Nt=18;Ng=Nv+Nt
allocate(chi(1:Ne,1:Ng))
!!!!vertex data!!!!
 chi(:,1)=(/1,0,1,1,0,0,0,0,1,0,0,0,0,0,0,0,0,0,0,0,0,1,1,0,0,0,0/)
 chi(:,2)=(/1,1,0,0,1,1,0,0,0,0,0,0,0,0,0,0,0,0,0,0,0,0,0,1,1,0,0/)
 chi(:,3)=(/0,1,1,0,0,0,1,1,0,0,0,0,0,0,0,0,0,0,0,0,0,0,0,0,0,1,1/)
 chi(:,4)=(/0,0,0,1,1,0,0,0,0,1,0,1,1,0,0,0,0,1,0,0,0,0,0,0,0,0,0/)
 chi(:,5)=(/0,0,0,0,0,1,1,0,0,1,1,0,0,1,1,0,0,0,0,0,0,0,0,0,0,0,0/)
 chi(:,6)=(/0,0,0,0,0,0,0,1,1,0,1,1,0,0,0,1,1,0,0,0,0,0,0,0,0,0,0/)
 chi(:,7)=(/0,0,0,0,0,0,0,0,0,0,0,0,1,1,0,0,0,0,1,0,1,1,0,0,0,0,1/)
 chi(:,8)=(/0,0,0,0,0,0,0,0,0,0,0,0,0,0,1,1,0,0,1,1,0,0,1,1,0,0,0/)
 chi(:,9)=(/0,0,0,0,0,0,0,0,0,0,0,0,0,0,0,0,1,1,0,1,1,0,0,0,1,1,0/)
!!!!triangle data!!!!
chi(:,10)=(/1,0,0,1,1,0,0,0,0,0,0,0,0,0,0,0,0,0,0,0,0,0,0,0,0,0,0/)
chi(:,11)=(/0,0,0,0,1,1,0,0,0,1,0,0,0,0,0,0,0,0,0,0,0,0,0,0,0,0,0/)
chi(:,12)=(/0,1,0,0,0,1,1,0,0,0,0,0,0,0,0,0,0,0,0,0,0,0,0,0,0,0,0/)
chi(:,13)=(/0,0,0,0,0,0,1,1,0,0,1,0,0,0,0,0,0,0,0,0,0,0,0,0,0,0,0/)
chi(:,14)=(/0,0,1,0,0,0,0,1,1,0,0,0,0,0,0,0,0,0,0,0,0,0,0,0,0,0,0/)
chi(:,15)=(/0,0,0,1,0,0,0,0,1,0,0,1,0,0,0,0,0,0,0,0,0,0,0,0,0,0,0/)
chi(:,16)=(/0,0,0,0,0,0,0,0,0,1,0,0,1,1,0,0,0,0,0,0,0,0,0,0,0,0,0/)
chi(:,17)=(/0,0,0,0,0,0,0,0,0,0,0,0,0,1,1,0,0,0,1,0,0,0,0,0,0,0,0/)
chi(:,18)=(/0,0,0,0,0,0,0,0,0,0,1,0,0,0,1,1,0,0,0,0,0,0,0,0,0,0,0/)
chi(:,19)=(/0,0,0,0,0,0,0,0,0,0,0,0,0,0,0,1,1,0,0,1,0,0,0,0,0,0,0/)
chi(:,20)=(/0,0,0,0,0,0,0,0,0,0,0,1,0,0,0,0,1,1,0,0,0,0,0,0,0,0,0/)
chi(:,21)=(/0,0,0,0,0,0,0,0,0,0,0,0,1,0,0,0,0,1,0,0,1,0,0,0,0,0,0/)
chi(:,22)=(/0,0,0,0,0,0,0,0,0,0,0,0,0,0,0,0,0,0,1,0,0,1,1,0,0,0,0/)
chi(:,23)=(/1,0,0,0,0,0,0,0,0,0,0,0,0,0,0,0,0,0,0,0,0,0,1,1,0,0,0/)
chi(:,24)=(/0,0,0,0,0,0,0,0,0,0,0,0,0,0,0,0,0,0,0,1,0,0,0,1,1,0,0/)
chi(:,25)=(/0,1,0,0,0,0,0,0,0,0,0,0,0,0,0,0,0,0,0,0,0,0,0,0,1,1,0/)
chi(:,26)=(/0,0,0,0,0,0,0,0,0,0,0,0,0,0,0,0,0,0,0,0,1,0,0,0,0,1,1/)
chi(:,27)=(/0,0,1,0,0,0,0,0,0,0,0,0,0,0,0,0,0,0,0,0,0,1,0,0,0,0,1/)
\end{lstlisting}
In this case, the Hilbert space dimension is $2^{27}$ and, by using \ref{Script:FullCode2}, we were able to resolve the lowest five eigenvalues and the corresponding eigenvectors. With the stop constant fixed at $\epsilon=10^{-4}$, the iteration exited after 45 steps and returned the following eigenvalues: -26.9999999089124, -26.9990609390985, -26.9979615347183, -26.9838728924651 and -22.9999846549236. With same input, the source code \ref{Script:FullCode3} was executed on a laptop and the output was virtually identical.

\vspace{0.2cm}

The triangulation of the 1-torus shown in Fig.~\ref{Fig:Torus}(c) is obtained by pinching the icosahedron at the two shaded triangles. This is a particularly interesting way to pass from genus 0 to genus 1 triangulations because pinching at different pairs of triangles enables one to create a class of genus 1 triangulations that have the same number of edges, hence toric models with same Hilbert space dimension. An interesting question we want to answer in the future is if the toric models over such triangulations can be homotopically connected without closing the spectral gap. In particular, we are interested to determine if the two pinched triangles can be adiabatically exchanged and what kind of braid matrix results from this action.

\vspace{0.2cm}

The input file corresponding to Fig.~\ref{Fig:Torus}(c) is supplied below.
 
 \vspace{0.2cm}
 
\begin{lstlisting}[caption={\ },label={Script:CTorus}]
Nv=9;Ne=27;Nt=18;Ng=Nv+Nt
allocate(chi(1:Ne,1:Ng))
!!!!vertex data!!!!
 chi(:,1)=(/1,1,1,1,1,0,0,0,0,0,0,0,0,0,0,0,0,0,0,0,0,0,0,0,1,1,1/)
 chi(:,2)=(/1,0,0,0,0,1,0,0,0,1,1,1,0,0,0,0,0,1,1,0,0,0,0,1,0,0,0/)
 chi(:,3)=(/0,1,0,0,0,1,1,0,0,0,1,0,1,1,0,0,0,0,0,1,1,0,0,0,0,0,0/)
 chi(:,4)=(/0,0,1,0,0,0,1,1,0,0,0,0,0,0,1,1,0,0,0,0,0,0,0,0,0,0,0/)
 chi(:,5)=(/0,0,0,1,0,0,0,1,1,0,0,0,0,0,0,0,1,1,0,0,0,0,0,0,0,0,0/)
 chi(:,6)=(/0,0,0,0,1,0,0,0,1,1,0,0,0,0,0,0,0,0,1,1,0,0,0,0,0,0,0/)
 chi(:,7)=(/0,0,0,0,0,0,0,0,0,0,0,1,1,0,0,0,0,0,0,0,1,1,0,0,1,0,0/)
 chi(:,8)=(/0,0,0,0,0,0,0,0,0,0,0,0,0,1,1,0,0,0,0,0,0,1,1,0,0,1,0/)
 chi(:,9)=(/0,0,0,0,0,0,0,0,0,0,0,0,0,0,0,1,1,0,0,0,0,0,1,1,0,0,1/)
!!!!triangle data!!!!\\
chi(:,10)=(/0,1,1,0,0,0,1,0,0,0,0,0,0,0,0,0,0,0,0,0,0,0,0,0,0,0,0/)
chi(:,11)=(/0,0,1,1,0,0,0,1,0,0,0,0,0,0,0,0,0,0,0,0,0,0,0,0,0,0,0/)
chi(:,12)=(/0,0,0,1,1,0,0,0,1,0,0,0,0,0,0,0,0,0,0,0,0,0,0,0,0,0,0/)
chi(:,13)=(/1,0,0,0,1,0,0,0,0,1,0,0,0,0,0,0,0,0,0,0,0,0,0,0,0,0,0/)
chi(:,14)=(/0,0,0,0,0,0,0,0,0,0,1,1,0,0,0,0,0,0,0,0,1,0,0,0,0,0,0/)
chi(:,15)=(/0,0,0,0,0,1,0,0,0,0,0,1,1,0,0,0,0,0,0,0,0,0,0,0,0,0,0/)
chi(:,16)=(/0,0,0,0,0,0,0,0,0,0,0,0,1,1,0,0,0,0,0,0,0,1,0,0,0,0,0/)
chi(:,17)=(/0,0,0,0,0,0,1,0,0,0,0,0,0,1,1,0,0,0,0,0,0,0,0,0,0,0,0/)
chi(:,18)=(/0,0,0,0,0,0,0,0,0,0,0,0,0,0,1,1,0,0,0,0,0,0,1,0,0,0,0/)
chi(:,19)=(/0,0,0,0,0,0,0,1,0,0,0,0,0,0,0,1,1,0,0,0,0,0,0,0,0,0,0/)
chi(:,20)=(/0,0,0,0,0,0,0,0,0,0,0,0,0,0,0,0,1,1,0,0,0,0,0,1,0,0,0/)
chi(:,21)=(/0,0,0,0,0,0,0,0,1,0,0,0,0,0,0,0,0,1,1,0,0,0,0,0,0,0,0/)
chi(:,22)=(/0,0,0,0,0,1,0,0,0,0,0,0,0,0,0,0,0,0,1,1,0,0,0,0,0,0,0/)
chi(:,23)=(/0,0,0,0,0,0,0,0,0,1,1,0,0,0,0,0,0,0,0,1,0,0,0,0,0,0,0/)
chi(:,24)=(/0,1,0,0,0,0,0,0,0,0,0,0,0,0,0,0,0,0,0,0,1,0,0,0,1,0,0/)
chi(:,25)=(/0,0,0,0,0,0,0,0,0,0,0,0,0,0,0,0,0,0,0,0,0,1,0,0,1,1,0/)
chi(:,26)=(/0,0,0,0,0,0,0,0,0,0,0,0,0,0,0,0,0,0,0,0,0,0,1,0,0,1,1/)
chi(:,27)=(/1,0,0,0,0,0,0,0,0,0,0,0,0,0,0,0,0,0,0,0,0,0,0,1,0,0,1/)
\end{lstlisting}

\vspace{0.2cm}

The Hilbert space dimension is again $2^{27}$ and, by using \ref{Script:FullCode2}, we were able to resolve the lowest five eigenvalues and the corresponding eigenvectors. With the stop constant fixed at $\epsilon=10^{-4}$, the iteration exited after 45 steps and returned the following eigenvalues: -26.9999993796453, -26.9994564165789, -26.9988516614014, -26.9924450489185 and -22.9999850957486. With same input, the source code \ref{Script:FullCode3} was executed on a laptop and the output was virtually identical.

\subsection{Orientable triangulations of genus 2}

\begin{figure}[h]
	\centering
	\includegraphics[width=\linewidth]{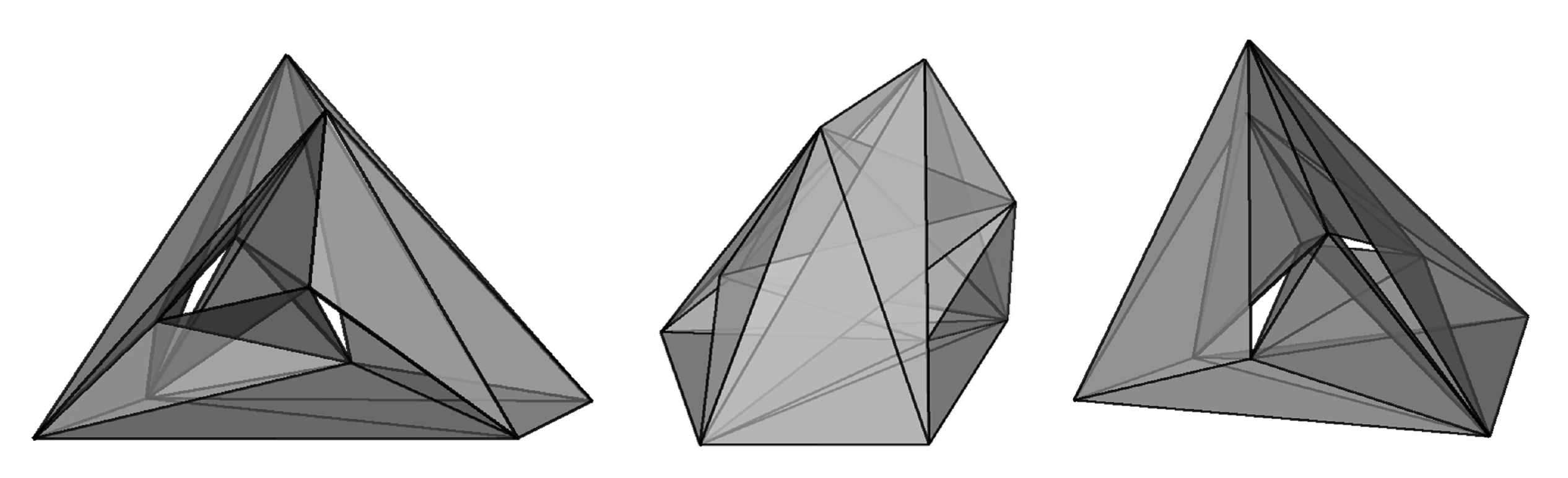}
	\caption{Views of the minimal Triangulations of the double torus.} 
	\label{Fig:DoubleTorus}
\end{figure}

The 3-dimensional rendering of the minimal triangulation of the 2-torus is shown in Fig.~\ref{Fig:DoubleTorus} and its corresponding input file is supplied below.

\vspace{0.2cm}

\begin{lstlisting}[basicstyle=\linespread{1.0}\ttfamily\tiny,caption={\ },label={Script:DoubleTorus}]
Nv=10;Ne=36;Nt=24;Ng=Nv+Nt 
allocate(chi(1:Ne,1:Ng)) 
!!!!vertex data!!!! 
 chi(:,1)=(/1,1,1,1,1,1,1,1,1,0,0,0,0,0,0,0,0,0,0,0,0,0,0,0,0,0,0,0,0,0,0,0,0,0,0,0/)
 chi(:,2)=(/1,0,0,0,0,0,0,0,0,1,1,1,1,1,1,0,0,0,0,0,0,0,0,0,0,0,0,0,0,0,0,0,0,0,0,0/)
 chi(:,3)=(/0,1,0,0,0,0,0,0,0,1,0,0,0,0,0,1,1,1,1,1,1,0,0,0,0,0,0,0,0,0,0,0,0,0,0,0/)
 chi(:,4)=(/0,0,1,0,0,0,0,0,0,0,1,0,0,0,0,0,0,0,0,0,0,1,1,1,1,1,1,0,0,0,0,0,0,0,0,0/)
 chi(:,5)=(/0,0,0,1,0,0,0,0,0,0,0,0,0,0,0,1,0,0,0,0,0,1,0,0,0,0,0,1,1,0,0,0,0,0,0,0/)
 chi(:,6)=(/0,0,0,0,1,0,0,0,0,0,0,1,0,0,0,0,1,0,0,0,0,0,1,0,0,0,0,0,0,1,1,0,0,0,0,0/)
 chi(:,7)=(/0,0,0,0,0,1,0,0,0,0,0,0,1,0,0,0,0,1,0,0,0,0,0,1,0,0,0,1,0,1,0,1,1,0,0,0/)
 chi(:,8)=(/0,0,0,0,0,0,1,0,0,0,0,0,0,1,0,0,0,0,1,0,0,0,0,0,1,0,0,0,0,0,1,0,0,1,1,0/)
 chi(:,9)=(/0,0,0,0,0,0,0,1,0,0,0,0,0,0,0,0,0,0,0,1,0,0,0,0,0,1,0,0,0,0,0,1,0,1,0,1/)
chi(:,10)=(/0,0,0,0,0,0,0,0,1,0,0,0,0,0,1,0,0,0,0,0,1,0,0,0,0,0,1,0,1,0,0,0,1,0,1,1/)
!!!!triangle data!!!!
chi(:,11)=(/1,1,0,0,0,0,0,0,0,1,0,0,0,0,0,0,0,0,0,0,0,0,0,0,0,0,0,0,0,0,0,0,0,0,0,0/)
chi(:,12)=(/1,0,1,0,0,0,0,0,0,0,1,0,0,0,0,0,0,0,0,0,0,0,0,0,0,0,0,0,0,0,0,0,0,0,0,0/)
chi(:,13)=(/0,1,0,1,0,0,0,0,0,0,0,0,0,0,0,1,0,0,0,0,0,0,0,0,0,0,0,0,0,0,0,0,0,0,0,0/)
chi(:,14)=(/0,0,1,0,1,0,0,0,0,0,0,0,0,0,0,0,0,0,0,0,0,0,1,0,0,0,0,0,0,0,0,0,0,0,0,0/)
chi(:,15)=(/0,0,0,1,0,1,0,0,0,0,0,0,0,0,0,0,0,0,0,0,0,0,0,0,0,0,0,1,0,0,0,0,0,0,0,0/)
chi(:,16)=(/0,0,0,0,1,0,1,0,0,0,0,0,0,0,0,0,0,0,0,0,0,0,0,0,0,0,0,0,0,0,1,0,0,0,0,0/)
chi(:,17)=(/0,0,0,0,0,1,0,1,0,0,0,0,0,0,0,0,0,0,0,0,0,0,0,0,0,0,0,0,0,0,0,1,0,0,0,0/)
chi(:,18)=(/0,0,0,0,0,0,1,0,1,0,0,0,0,0,0,0,0,0,0,0,0,0,0,0,0,0,0,0,0,0,0,0,0,0,1,0/)
chi(:,19)=(/0,0,0,0,0,0,0,1,1,0,0,0,0,0,0,0,0,0,0,0,0,0,0,0,0,0,0,0,0,0,0,0,0,0,0,1/)
chi(:,20)=(/0,0,0,0,0,0,0,0,0,1,0,1,0,0,0,0,1,0,0,0,0,0,0,0,0,0,0,0,0,0,0,0,0,0,0,0/)
chi(:,21)=(/0,0,0,0,0,0,0,0,0,0,1,0,0,1,0,0,0,0,0,0,0,0,0,0,1,0,0,0,0,0,0,0,0,0,0,0/)
chi(:,22)=(/0,0,0,0,0,0,0,0,0,0,0,1,1,0,0,0,0,0,0,0,0,0,0,0,0,0,0,0,0,1,0,0,0,0,0,0/)
chi(:,23)=(/0,0,0,0,0,0,0,0,0,0,0,0,1,0,1,0,0,0,0,0,0,0,0,0,0,0,0,0,0,0,0,0,1,0,0,0/)
chi(:,24)=(/0,0,0,0,0,0,0,0,0,0,0,0,0,1,1,0,0,0,0,0,0,0,0,0,0,0,0,0,0,0,0,0,0,0,1,0/)
chi(:,25)=(/0,0,0,0,0,0,0,0,0,0,0,0,0,0,0,1,0,0,0,0,1,0,0,0,0,0,0,0,1,0,0,0,0,0,0,0/)
chi(:,26)=(/0,0,0,0,0,0,0,0,0,0,0,0,0,0,0,0,1,0,1,0,0,0,0,0,0,0,0,0,0,0,1,0,0,0,0,0/)
chi(:,27)=(/0,0,0,0,0,0,0,0,0,0,0,0,0,0,0,0,0,1,0,1,0,0,0,0,0,0,0,0,0,0,0,1,0,0,0,0/)
chi(:,28)=(/0,0,0,0,0,0,0,0,0,0,0,0,0,0,0,0,0,1,0,0,1,0,0,0,0,0,0,0,0,0,0,0,1,0,0,0/)
chi(:,29)=(/0,0,0,0,0,0,0,0,0,0,0,0,0,0,0,0,0,0,1,1,0,0,0,0,0,0,0,0,0,0,0,0,0,1,0,0/)
chi(:,30)=(/0,0,0,0,0,0,0,0,0,0,0,0,0,0,0,0,0,0,0,0,0,1,0,1,0,0,0,1,0,0,0,0,0,0,0,0/)
chi(:,31)=(/0,0,0,0,0,0,0,0,0,0,0,0,0,0,0,0,0,0,0,0,0,1,0,0,0,0,1,0,1,0,0,0,0,0,0,0/)
chi(:,32)=(/0,0,0,0,0,0,0,0,0,0,0,0,0,0,0,0,0,0,0,0,0,0,1,1,0,0,0,0,0,1,0,0,0,0,0,0/)
chi(:,33)=(/0,0,0,0,0,0,0,0,0,0,0,0,0,0,0,0,0,0,0,0,0,0,0,0,1,1,0,0,0,0,0,0,0,1,0,0/)
chi(:,34)=(/0,0,0,0,0,0,0,0,0,0,0,0,0,0,0,0,0,0,0,0,0,0,0,0,0,1,1,0,0,0,0,0,0,0,0,1/)
\end{lstlisting}

\vspace{0.2cm}

The Hilbert space dimension is $2^{36}$, which was too large for our current computational resources.

\section{Final Remarks and Conclusions}

In our opinion, one of the important achievements of this work is enabling laptop simulations of spin systems that are as complex as the toric model deployed on triangulations of genus 1. The implication is that researchers as well as scholars studying the field can explore these models numerically even if they are not affiliated to a computational laboratory. In particular, these simulations can now be performed in classrooms. For this reason, we want to report a few facts about the performance of our code on Microsoft's Surface Laptop 2 with the following cpu configuration:
\begin{lstlisting}
*-cpu
   ${\rm product}$: Intel(R) Core(TM) i5-8350U CPU @ 1.70GHz
   vendor: Intel Corp.
   capacity: 1896MHz
   width: 64 bits
\end{lstlisting}
and memory configuration:
\begin{lstlisting}
        total             
Mem:    16700476     
Swap:   31094908     
\end{lstlisting}
The fortran script \ref{Script:FullCode3} was compiled with the first 2020 Linux release of Intel$^{\mbox{\textregistered}}$ Parallel Studio XE.\footnote{Installed under Windows Subsystem for Linux.} We should mention that the cpu mentioned above supports 8 threads and the laptop supports Intel$^{\mbox{\textregistered}}$ Rapid Storage Technology, which enable efficient use of swap memory. The latter is a feature that is absolutely essential for our applications.

\vspace{0.2cm}

For all inputs, the main execution tasks, {\it i.e.} the two large loops over $\Gamma$, were fully parallelized on all eight available threads, as the report below shows:
\begin{lstlisting}[basicstyle=\linespread{1.0}\ttfamily\scriptsize]
%Cpu0  : 96.4 us,  3.6 sy,  0.0 ni,  0.0 id,  0.0 wa,  0.0 hi,  0.0 si
%Cpu1  : 96.9 us,  2.5 sy,  0.0 ni,  0.5 id,  0.0 wa,  0.0 hi,  0.0 si
%Cpu2  : 99.0 us,  1.0 sy,  0.0 ni,  0.0 id,  0.0 wa,  0.0 hi,  0.0 si
%Cpu3  : 99.5 us,  0.5 sy,  0.0 ni,  0.0 id,  0.0 wa,  0.0 hi,  0.0 si
%Cpu4  : 78.6 us, 21.2 sy,  0.0 ni,  0.0 id,  0.0 wa,  0.3 hi,  0.0 si
%Cpu5  : 91.6 us,  8.4 sy,  0.0 ni,  0.0 id,  0.0 wa,  0.0 hi,  0.0 si
%Cpu6  : 97.2 us,  2.8 sy,  0.0 ni,  0.0 id,  0.0 wa,  0.0 hi,  0.0 si
%Cpu7  : 99.5 us,  0.5 sy,  0.0 ni,  0.0 id,  0.0 wa,  0.0 hi,  0.0 si
\end{lstlisting}

\vspace{0.2cm}

With the input \ref{Script:ATorus} for which the Hilbert space dimension is $2^{21}$, the memory management was as follows:
\begin{lstlisting}
        total       used       free        shared   buff      available
Mem:    16700476    5739976    10731148    17720    229352    10826768
Swap:   31094908    1355540    29739368
\end{lstlisting}
while the execution time was:
\begin{lstlisting}
real    8m46.906s
user    54m36.500s
sys     0m9.094s
\end{lstlisting}
Hence, it takes a mere 9 minutes to resolve the 4-fold degenerate ground state and one excited state. 

\vspace{0.2cm}

With the input \ref{Script:BTorus} for which the Hilbert space dimension is $2^{27}$, the memory management was as follows:
\begin{lstlisting}[basicstyle=\linespread{1.0}\ttfamily\scriptsize]
        total      used        free      shared   buff      available
Mem:    16700476   15690580    780544    17720    229352    876164
Swap:   31094908   14725908    16369000
\end{lstlisting}
while the execution time was:
\begin{lstlisting}[basicstyle=\linespread{1.0}\ttfamily\scriptsize]
real    1546m17.338s
user    6008m48.281s
sys     332m16.125s
\end{lstlisting}
Hence, it takes about 26 hours or a little bit over one day to resolve the 4-fold degenerate ground state and one excited state. Let us specify that the stop parameter was fixed as $\epsilon = 10^{-4}$ and that the iteration exited the 45-th step and returned the following eigenvalues: -26.9999999402053, -26.9991332763039, -26.9979230003306, -26.9906901387635, and -22.9999855049633.

\vspace{0.2cm}

As it becomes apparent from our applications, the Hilbert space dimensions of the toric models increase rapidly with the complexity and the size of triangulation. Note that we have only dealt with the minimal version of the toric code, which builds on the quantum double of the abelian group $\ZM_2$. As it is well-known \cite{KitaevAOP2003}, the model extends naturally to generic finite groups, both abelian and non-abelian. To simulate such general toric models, one has to increase the dimension $m$ of the local Hilbert spaces and this will add to the computational complexity. Nevertheless, the source codes and the input files we supplied can be straightforwardly modified to cover these extensions. This one direction which we want to explore in the near future. Another direction is simulating the anyon states and their braidings.

\vspace{0.2cm}

The source code can be also adapted for more traditional lattice spin systems, which generally involve structured Hamiltonians. Also, quite often, these models have one or more conserved quantities, such as the global magnetization, which can reduce the computational complexity. Applications to this class of systems will be considered in the future.

\section*{Acknowledgements} This work was supported by an award from the W.M. Keck Foundation and by NSF through the grant DMR-1823800.

\appendix
\section{} 
\label{App:1}
 
This is the Fortran source code which computes and diagonalizes the full Hamiltonian of the toric code deployed on the triangulation \ref{Fig:Sphere}(b). The input file is Octahedron.f90, which contains an identical copy of Block~\ref{Script:Octahedron}.

 \begin{lstlisting}[language=Fortran,caption={\ },label={Script:FullCode1}]
      implicit none
!     ************************************
!     triangulation input data
!     ************************************
      integer Nv,Ne,Nt,Ng  !nr of vertices,edges,triangles,subsets
      integer,allocatable :: chi(:,:)
!     ************************************
!     structure data
!     ************************************
      integer ga,Qmax
      integer,allocatable :: Q(:)
      integer,allocatable :: nn(:,:)
!     ************************************
!     local operator data
!     ************************************
      integer m  !dim of local Hilbert
      parameter(m=2)
      integer,allocatable :: K(:,:)
      integer,allocatable :: Nk(:)
      integer,allocatable :: g1(:,:,:)
      integer,allocatable :: g2(:,:,:)
      double precision,allocatable :: AA(:,:,:)
      double precision prodA
!     ************************************
!     global operator data
!     ************************************
      integer dimH  !global Hilbert dimension
      integer Ind1,Ind2  !Hamilton indices
      double precision,allocatable :: Ham(:,:)
      double precision,allocatable :: eig(:)
      double precision,allocatable :: wr(:)
      integer inf
!     ************************************
      integer,allocatable :: kk(:),ii(:)  !important indices
      integer a,b,r,s
      integer proc,w,z
      integer e,ct,expo
!     ************************************
!     triangulation/Hamiltonian input data
!     ************************************
      include 'Octahedron.f90'
!     ************************************
!     structure data Q,nn
!     ************************************
      allocate(Q(1:Ng))
      do ga=1,Ng
       Q(ga)=sum(chi(:,ga))+1
      end do
      Qmax=maxval(Q(:))
      allocate(nn(1:Qmax,1:Ng))
      nn(:,:)=0
      do ga=1,Ng
       r=1;ct=0
       do e=1,Ne
        if(chi(e,ga).eq.0) then
         nn(r,ga)=nn(r,ga)+1
        else
         r=r+1
        end if
       end do
      end do
!     ************************************
!     input data for local operators
!     ************************************
      allocate(K(1:Qmax-1,1:Ng))
      allocate(g1(0:m**2-1,1:Qmax-1,1:Ng))
      allocate(g2(0:m**2-1,1:Qmax-1,1:Ng))
      allocate(AA(0:m**2-1,1:Qmax-1,1:Ng))
      K=0;g1=0;g2=0;AA=0
      do ga=1,Ng
       do r=1,Q(ga)-1   
        K(r,ga)=2
        if (ga.le.Nv) then 
         g1(0,r,ga)=1-1; g2(0,r,ga)=1-1; AA(0,r,ga)=1
         g1(1,r,ga)=2-1; g2(1,r,ga)=2-1; AA(1,r,ga)=-1
        else
         g1(0,r,ga)=1-1; g2(0,r,ga)=2-1; AA(0,r,ga)=1
         g1(1,r,ga)=2-1; g2(1,r,ga)=1-1; AA(1,r,ga)=1
        end if
       end do
      end do
      allocate(Nk(1:Ng))
      Nk=1
      do ga=1,Ng
       do r=1,Q(ga)-1
        Nk(ga)=Nk(ga)*K(r,ga)
       end do
      end do
!     ************************************
!     generates the Hamiltonian
!     ************************************
      dimH=m**Ne
      allocate(Ham(1:dimH,1:dimH))
      allocate(eig(1:dimH))
      allocate(wr(12*dimH))
      allocate(kk(1:Qmax-1))
      allocate(ii(1:Qmax))
      Ham=0d0
      do ga=1,Ng
!      ***********************************
!      sum over alpha begins
!      ***********************************
       do a=0,Nk(ga)-1
!       ***********************************
!       generates the kk indices
!       ***********************************
        proc=0;w=1;kk=0
        do r=1,Q(ga)-1
         z=K(r,ga)
         kk(r)=modulo((a-proc)/w,z)
         proc=proc+kk(r)*w
         w=w*z
        end do
!       ***********************************
!       computes the product of A's
!       ***********************************
        prodA=1d0
        do r=1,Q(ga)-1
         prodA=prodA*AA(kk(r),r,ga)
        end do
!       ***********************************
!       sum over beta begins
!       ***********************************
        do b=0,m**(Ne-Q(ga)+1)-1
!        **********************************
!        generates the ii indices
!        **********************************
         proc=0;w=1;ii=0
         do r=1,Q(ga)
          if(nn(r,ga).ne.0) then
           z=m**nn(r,ga)
           ii(r)=modulo((b-proc)/w,z) 
           proc=proc+ii(r)*w
           w=w*z
          end if
         end do
!        **********************************
!        populates the non-zero matrix elements
!        **********************************
         Ind1=ii(Q(ga))+1
         Ind2=ii(Q(ga))+1
         do r=1,Q(ga)-1
          expo=sum(nn(r+1:Q(ga),ga))+Q(ga)-r
          Ind1=Ind1+ii(r)*m**expo+g1(kk(r),r,ga)*m**(expo-1)
          Ind2=Ind2+ii(r)*m**expo+g2(kk(r),r,ga)*m**(expo-1) 
         end do 
         Ham(Ind1,Ind2)=Ham(Ind1,Ind2)-prodA
        end do 
       end do 
      end do  
!     ************************************
!     diagonalizes the global Hamiltonian
!     ************************************       
      call dsyev('n','l',dimH,Ham,dimH,eig,wr,12*dimH,inf)
      open(11,file='eig.txt')
      do a=1,dimH
       write(11,*) a,eig(a)
      end do
      end
\end{lstlisting}
 
\section{}
\label{App:2}

This is the Fortran source code which computes the first $p=5$ eigenvalues and corresponding eigenvectors of the toric mode deployed on the triangulation \ref{Fig:Torus}(a). The input file is ATorus.f90, which contains an identical copy of Block~\ref{Script:ATorus}. Sections of the codes have been parallelized using simple OpenMP directives.

 \begin{lstlisting}[language=Fortran,caption={\ },label={Script:FullCode2}]
      implicit none
!     ************************************
!     triangulation input data
!     ************************************
      integer Nv,Ne,Nt,Ng  !nr of vertices,edges,triangles,subsets
      integer,allocatable :: chi(:,:)
!     ************************************
!     structure data
!     ************************************
      integer ga,Qmax
      integer,allocatable :: Q(:)
      integer,allocatable :: nn(:,:)
!     ************************************
!     local operator data
!     ************************************
      integer m  !dim of local Hilbert
      parameter(m=2)
      integer,allocatable :: K(:,:)
      integer,allocatable :: Nk(:)
      integer,allocatable :: g1(:,:,:)
      integer,allocatable :: g2(:,:,:)
      double precision,allocatable :: AA(:,:,:)
      double complex prodA
!     ************************************
!     global operator data
!     ************************************
      integer (kind=8) dimH  !global Hilbert dimension
      integer (kind=8) Ind1,Ind2
      integer p,iter
      parameter(p=5,iter=60)
      double precision eps,cut
      parameter(eps=0.0001d0,cut=exp(30d0))
      double precision,allocatable :: GH(:,:,:)
      double precision,allocatable :: PH(:,:)
      double precision,allocatable :: psi(:,:),opsi(:,:)
      double precision,allocatable :: gpsi(:,:,:)
      double precision,allocatable :: eig(:)
      double precision,allocatable :: SS(:,:),SSS(:,:)
      double precision,allocatable :: wr(:)
      double precision harvest,ortho,val
      integer inf
!     ************************************
      integer,allocatable :: kk(:,:),ii(:,:)  !important indices
      integer a,b,r,s,u,v,j,i
      integer proc,w,z
      integer e,ct,expo,dd
!     ************************************
!     triangulation/Hamiltonian input data
!     ************************************
      include 'ATorus.f90'
!     ************************************
!     structure data Q,nn
!     ************************************
      allocate(Q(1:Ng))
      do concurrent(ga=1:Ng)
       Q(ga)=sum(chi(:,ga))+1
      end do
      Qmax=maxval(Q(:))
      allocate(nn(1:Qmax,1:Ng))
      nn(:,:)=0
      do concurrent (ga=1:Ng)
       r=1;ct=0
       do e=1,Ne
        if(chi(e,ga).eq.0) then
         nn(r,ga)=nn(r,ga)+1
        else
         r=r+1
        end if
       end do
      end do
!     ************************************
!     input data for local operators
!     ************************************
      allocate(K(1:Qmax-1,1:Ng))
      allocate(g1(0:m**2-1,1:Qmax-1,1:Ng))
      allocate(g2(0:m**2-1,1:Qmax-1,1:Ng))
      allocate(AA(0:m**2-1,1:Qmax-1,1:Ng))
      K=0;g1=0;g2=0;AA=0
      do ga=1,Ng
       do r=1,Q(ga)-1
        K(r,ga)=2
        if (ga.le.Nv) then 
         g1(0,r,ga)=1-1; g2(0,r,ga)=1-1; AA(0,r,ga)=1
         g1(1,r,ga)=2-1; g2(1,r,ga)=2-1; AA(1,r,ga)=-1
        else
         g1(0,r,ga)=1-1; g2(0,r,ga)=2-1; AA(0,r,ga)=1
         g1(1,r,ga)=2-1; g2(1,r,ga)=1-1; AA(1,r,ga)=1
        end if
       end do
      end do
      allocate(Nk(1:Ng))
      Nk=1
      do ga=1,Ng
       do r=1,Q(ga)-1
        Nk(ga)=Nk(ga)*K(r,ga)
       end do
      end do
!     ************************************
!     generates the projected Hamiltonian
!     ************************************
      dimH=m**Ne
      allocate(psi(1:dimH,1:2*p),opsi(1:dimH,1:2*p))
      allocate(gpsi(1:dimH,1:p,1:Ng))
      allocate(SS(1:2*p,1:2*p),SSS(1:2*p,1:2*p))
      allocate(PH(1:2*p,1:2*p),eig(1:2*p))
      allocate(GH(1:2*p,1:2*p,1:Ng))
      allocate(wr(12*2*p))
      allocate(kk(1:Qmax-1,1:Ng))
      allocate(ii(1:Qmax,1:Ng))
!     ************************************
!     intitiate psi's
!     ************************************
      psi=0d0
      call random_seed()
      do u=1,2*p
       do j=1,dimH
        call random_number(harvest)
        psi(j,u)=harvest  !generate 2p random psi
       end do
      end do
!     ************************************
!     start iterative process
!     ************************************
      do j=1,iter  
!      ***********************************
!      ortho-normalization of psi's
!      ***********************************
       SS=0d0
       ${\rm !\$OMP \ parallel \ do}$
       do v=1,2*p
        do u=v,2*p
         SS(u,v)=sum(psi(:,u)*psi(:,v))  !computes overlap matrix
        end do
       end do     
       ${\rm !\$OMP \ end \ parallel \ do}$
       call dsyev('v','l',2*p,SS,2*p,eig,wr,12*2*p,inf)
       write(*,*) '**************************'
       write(*,*) 'spectrum of overlap matrix'
       write(*,*)  eig(1:p)
       write(*,*) '**************************'
       if(eig(p).lt.eps) go to 200
       SSS=0d0
       ${\rm !\$OMP \ parallel \ do}$
       do v=1,2*p
        do u=1,2*p
         val=min(1d0/dsqrt(eig(u)),cut)
         SSS(:,v)=SSS(:,v)+val*SS(:,u)*SS(v,u) !computes S^{-1/2}
        end do
       end do
       ${\rm !\$OMP \ end \ parallel \ do}$
       opsi=0d0  !orthogonal psi's
       ${\rm !\$OMP \ parallel \ do}$
       do u=1,2*p
        do v=1,2*p
         opsi(:,u)=opsi(:,u)+SSS(u,v)*psi(:,v)
        end do
       end do
       ${\rm !\$OMP \ end \ parallel \ do}$
!      *************************************
!      Generated PH using opsi's
!      *************************************
       PH=0d0
       GH=0d0
       ${\rm !\$OMP \ parallel \ do}$
       do ga=1,Ng
!       ************************************
!       sum over alpha begins
!       ************************************
        do a=0,Nk(ga)-1
!        ***********************************
!        generates the kk indices
!        ***********************************
         proc=0;w=1;kk(:,ga)=0
         do r=1,Q(ga)-1
          z=K(r,ga)
          kk(r,ga)=modulo((a-proc)/w,z)
          proc=proc+kk(r,ga)*w
          w=w*z
         end do
!        ***********************************
!        computes the product of A's
!        ***********************************
         prodA=1d0
         do r=1,Q(ga)-1
          prodA=prodA*AA(kk(r,ga),r,ga)
         end do
!        ***********************************
!        sum over beta begins
!        ***********************************
         do b=0,m**(Ne-Q(ga)+1)-1
!         **********************************
!         generates the ii indices
!         **********************************
          proc=0;w=1;ii(:,ga)=0
          do r=1,Q(ga)
           if(nn(r,ga).ne.0) then
            z=m**nn(r,ga)
            ii(r,ga)=modulo((b-proc)/w,z) 
            proc=proc+ii(r,ga)*w
            w=w*z
           end if
          end do
!         **********************************
!         computes the relevant H indices
!         **********************************
          Ind1=ii(Q(ga),ga)+1
          Ind2=ii(Q(ga),ga)+1
          do r=1,Q(ga)-1
           expo=sum(nn(r+1:Q(ga),ga))+Q(ga)-r
           Ind1=Ind1+ii(r,ga)*m**expo+g1(kk(r,ga),r,ga)*m**(expo-1)
           Ind2=Ind2+ii(r,ga)*m**expo+g2(kk(r,ga),r,ga)*m**(expo-1) 
          end do
          do v=1,2*p
           GH(:,v,ga)=GH(:,v,ga)-opsi(Ind1,:)*prodA*opsi(Ind2,v)
          end do
         end do 
        end do 
       end do
       ${\rm !\$OMP \ end \ parallel \ do}$
       do ga=1,Ng
        PH=PH+GH(:,:,ga)
       end do  
!      *************************************
!      diagonalizes projected Hamiltonian
!      *************************************       
       call dsyev('v','l',2*p,PH,2*p,eig,wr,12*2*p,inf)
       write(*,*) 'current Ham eigenvalues'
       write(*,*) j,eig(1:p)
!      *************************************
!      generates the first p new psi's
!      *************************************
       psi=0d0;gpsi=0d0
       ${\rm !\$OMP \ parallel \ do}$
       do u=1,p
        do v=1,2*p
         psi(:,u)=psi(:,u)+PH(v,u)*opsi(:,v)
        end do
       end do
       ${\rm !\$OMP \ end \ parallel \ do}$
!      *************************************
!      the next p new psi's are generated
!      *************************************
       ${\rm !\$OMP \ parallel \ do}$
       do ga=1,Ng
!       ************************************
!       sum over alpha begins
!       ************************************
        do a=0,Nk(ga)-1
!        ***********************************
!        generates the kk indices
!        ***********************************
         proc=0;w=1;kk(:,ga)=0
         do r=1,Q(ga)-1
          z=K(r,ga)
          kk(r,ga)=modulo((a-proc)/w,z)
          proc=proc+kk(r,ga)*w
          w=w*z
         end do
!        ***********************************
!        computes the product of A's
!        ***********************************
         prodA=1d0
         do r=1,Q(ga)-1
          prodA=prodA*AA(kk(r,ga),r,ga)
         end do
!        ***********************************
!        sum over beta begins
!        ***********************************
         do b=0,m**(Ne-Q(ga)+1)-1
!         **********************************
!         generates the ii indices
!         **********************************
          proc=0;w=1;ii(:,ga)=0
          do r=1,Q(ga)
           if(nn(r,ga).ne.0) then
            z=m**nn(r,ga)
            ii(r,ga)=modulo((b-proc)/w,z) 
            proc=proc+ii(r,ga)*w
            w=w*z
           end if
          end do
!         **********************************
!         generates relevant indices
!         **********************************
          Ind1=ii(Q(ga),ga)+1
          Ind2=ii(Q(ga),ga)+1
          do r=1,Q(ga)-1
           expo=sum(nn(r+1:Q(ga),ga))+Q(ga)-r
           Ind1=Ind1+ii(r,ga)*m**expo+g1(kk(r,ga),r,ga)*m**(expo-1)
           Ind2=Ind2+ii(r,ga)*m**expo+g2(kk(r,ga),r,ga)*m**(expo-1) 
          end do 
          do u=1,p
           gpsi(Ind1,u,ga)=gpsi(Ind1,u,ga)-prodA*psi(Ind2,u)
          end do
         end do 
        end do 
       end do  
       ${\rm !\$OMP \ end \ parallel \ do}$
       ${\rm !\$OMP \ parallel \ do}$
       do u=p+1,2*p
        do ga=1,Ng
         psi(:,u)=psi(:,u)+gpsi(:,u-p,ga)
        end do
       end do
       ${\rm !\$OMP \ end \ parallel \ do}$
      end do
 200  continue
      end
\end{lstlisting} 

\section{}
\label{App:3}

This is a modified version of \ref{Script:FullCode3}, which computes the first $p=5$ eigenvalues and corresponding eigenvectors of the toric mode deployed on the triangulation \ref{Fig:Torus}(b). The input file is BTorus.f90, which contains an identical copy of Block~\ref{Script:BTorus}. Sections of the codes have been again parallelized using simple OpenMP directives. Compared with \ref{Script:FullCode2}, this script does not allocate the large arrays gpsi, which were used in \ref{Script:FullCode3} to make the tasks inside the loop over $\Gamma$ explicitly independent. 
 \begin{lstlisting}[language=Fortran,caption={\ },label={Script:FullCode3}]
      implicit none
!     ************************************
!     triangulation input data
!     ************************************
      integer Nv,Ne,Nt,Ng  !nr of vertices,edges,triangles,subsets
      integer, allocatable :: chi(:,:)
!     ************************************
!     structure data
!     ************************************
      integer ga,Qmax
      integer, allocatable :: Q(:)
      integer, allocatable :: nn(:,:)
!     ************************************
!     local operator data
!     ************************************
      integer m  !dim of local Hilbert
      parameter(m=2)
      integer, allocatable :: K(:,:)
      integer, allocatable :: Nk(:)
      integer, allocatable :: g1(:,:,:)
      integer, allocatable :: g2(:,:,:)
      double precision, allocatable :: AA(:,:,:)
      double complex prodA
!     ************************************
!     global operator data
!     ************************************
      integer (kind=8) dimH  !global Hilbert dimension
      integer (kind=8) Ind1,Ind2
      integer p,iter
      parameter(p=5,iter=60)
      double precision eps,cut
      parameter(eps=0.0001d0,cut=exp(30d0))
      double precision, allocatable :: GH(:,:,:)
      double precision, allocatable :: PH(:,:)
      double precision, allocatable :: psi(:,:),opsi(:,:)
      double precision,allocatable :: eig(:)
      double precision, allocatable :: SS(:,:),SSS(:,:)
      double precision,allocatable :: wr(:)
      double precision harvest,ortho,val
      integer inf
!     ************************************
      integer, allocatable :: kk(:,:),ii(:,:)  !important indices
      integer a,b,r,s,u,v,j,i
      integer proc,w,z
      integer e,ct,expo,dd
!     ************************************
!     triangulation/Hamiltonian input data
!     ************************************
      include 'BTorus.f90'
!     ************************************
!     structure data Q,nn
!     ************************************
      allocate(Q(1:Ng))
      do concurrent(ga=1:Ng)
       Q(ga)=sum(chi(:,ga))+1
      end do
      Qmax=maxval(Q(:))
      allocate(nn(1:Qmax,1:Ng))
      nn(:,:)=0
      do concurrent (ga=1:Ng)
       r=1;ct=0
       do e=1,Ne
        if(chi(e,ga).eq.0) then
         nn(r,ga)=nn(r,ga)+1
        else
         r=r+1
        end if
       end do
      end do
!     ************************************
!     input data for local operators
!     ************************************
      allocate(K(1:Qmax-1,1:Ng))
      allocate(g1(0:m**2-1,1:Qmax-1,1:Ng))
      allocate(g2(0:m**2-1,1:Qmax-1,1:Ng))
      allocate(AA(0:m**2-1,1:Qmax-1,1:Ng))
      K=0;g1=0;g2=0;AA=0
      do ga=1,Ng
       do r=1,Q(ga)-1
        K(r,ga)=2
        if (ga.le.Nv) then 
         g1(0,r,ga)=1-1; g2(0,r,ga)=1-1; AA(0,r,ga)=1
         g1(1,r,ga)=2-1; g2(1,r,ga)=2-1; AA(1,r,ga)=-1
        else
         g1(0,r,ga)=1-1; g2(0,r,ga)=2-1; AA(0,r,ga)=1
         g1(1,r,ga)=2-1; g2(1,r,ga)=1-1; AA(1,r,ga)=1
        end if
       end do
      end do
      allocate(Nk(1:Ng))
      Nk=1
      do ga=1,Ng
       do r=1,Q(ga)-1
        Nk(ga)=Nk(ga)*K(r,ga)
       end do
      end do
!     ************************************
!     generates the projected Hamiltonian
!     ************************************
      dimH=m**Ne
      allocate(psi(1:dimH,1:2*p),opsi(1:dimH,1:2*p))
      allocate(SS(1:2*p,1:2*p),SSS(1:2*p,1:2*p))
      allocate(PH(1:2*p,1:2*p),eig(1:2*p))
      allocate(GH(1:2*p,1:2*p,1:Ng))
      allocate(wr(12*2*p))
      allocate(kk(1:Qmax-1,1:Ng))
      allocate(ii(1:Qmax,1:Ng))
!     ************************************
!     intitiate psi's
!     ************************************
      psi=0d0
      call random_seed()
      do u=1,2*p
       do j=1,dimH
        call random_number(harvest)
        psi(j,u)=harvest  !generate 2p random psi
       end do
      end do
!     ************************************
!     start iterative process
!     ************************************
      do j=1,iter  
!      ***********************************
!      ortho-normalization of psi's
!      ***********************************
       SS=0d0
       ${\rm !\$OMP \ parallel \ do}$
       do v=1,2*p
        do u=v,2*p
         SS(u,v)=sum(psi(:,u)*psi(:,v))  !computes overlap matrix
        end do
       end do     
       ${\rm !\$OMP \ end \ parallel \ do}$
       call dsyev('v','l',2*p,SS,2*p,eig,wr,12*2*p,inf)
       write(*,*) '**************************'
       write(*,*) 'spectrum of overlap matrix'
       write(*,*)  eig(1:p)
       write(*,*) '**************************'
       if(eig(p).lt.eps) go to 200
       SSS=0d0
       ${\rm !\$OMP \ parallel \ do}$
       do v=1,2*p
        do u=1,2*p
         val=min(1d0/dsqrt(eig(u)),cut)
         SSS(:,v)=SSS(:,v)+val*SS(:,u)*SS(v,u) !computes S^{-1/2}
        end do
       end do
       ${\rm !\$OMP \ end \ parallel \ do}$
       opsi=0d0
       ${\rm !\$OMP \ parallel \ do}$
       do u=1,2*p
        do v=1,2*p
         opsi(:,u)=opsi(:,u)+SSS(u,v)*psi(:,v)
        end do
       end do
       ${\rm !\$OMP \ end \ parallel \ do}$
!      ***********************************
!      Generated PH using opsi's
!      ***********************************
       PH=0d0
       GH=0d0
       ${\rm !\$OMP \ parallel \ do}$
       do ga=1,Ng
!       ***********************************
!       sum over alpha begins
!       ***********************************
        do a=0,Nk(ga)-1
!        **********************************
!        generates the kk indices
!        **********************************
         proc=0;w=1;kk(:,ga)=0
         do r=1,Q(ga)-1
          z=K(r,ga)
          kk(r,ga)=modulo((a-proc)/w,z)
          proc=proc+kk(r,ga)*w
          w=w*z
         end do
!        **********************************
!        computes the product of A's
!        **********************************
         prodA=1d0
         do r=1,Q(ga)-1
          prodA=prodA*AA(kk(r,ga),r,ga)
         end do
!        **********************************
!        sum over beta begins
!        **********************************
         do b=0,m**(Ne-Q(ga)+1)-1
!         *********************************
!         generates the ii indices
!         *********************************
          proc=0;w=1;ii(:,ga)=0
          do r=1,Q(ga)
           if(nn(r,ga).ne.0) then
            z=m**nn(r,ga)
            ii(r,ga)=modulo((b-proc)/w,z) 
            proc=proc+ii(r,ga)*w
            w=w*z
           end if
          end do
!         **********************************
!         computes the relevant H indices
!         **********************************
          Ind1=ii(Q(ga),ga)+1
          Ind2=ii(Q(ga),ga)+1
          do r=1,Q(ga)-1
           expo=sum(nn(r+1:Q(ga),ga))+Q(ga)-r
           Ind1=Ind1+ii(r,ga)*m**expo+g1(kk(r,ga),r,ga)*m**(expo-1)
           Ind2=Ind2+ii(r,ga)*m**expo+g2(kk(r,ga),r,ga)*m**(expo-1) 
          end do
          do v=1,2*p
            GH(:,v,ga)=GH(:,v,ga)-opsi(Ind1,:)*prodA*opsi(Ind2,v)
          end do
         end do 
        end do 
       end do
       ${\rm !\$OMP \ end \ parallel \ do}$
       do ga=1,Ng
        PH=PH+GH(:,:,ga)
       end do  
!      ***********************************
!      diagonalizes projected Hamiltonian
!      ***********************************     
       call dsyev('v','l',2*p,PH,2*p,eig,wr,12*2*p,inf)
       write(*,*) 'current Ham eigenvalues'
       write(*,*) j,eig(1:p)
!      ***********************************
!      generates the first p new psi's
!      ***********************************
       psi=0d0
       ${\rm !\$OMP \ parallel \ do}$
       do u=1,p
        do v=1,2*p
         psi(:,u)=psi(:,u)+PH(v,u)*opsi(:,v)
        end do
       end do
       ${\rm !\$OMP \ end \ parallel \ do}$
!      ************************************
!      the next p new psi's are generated
!      ************************************
       ${\rm !\$OMP \ parallel \ do}$
       do ga=1,Ng
!       ***********************************
!       sum over alpha begins
!       ***********************************
        do a=0,Nk(ga)-1
!        **********************************
!        generates the kk indices
!        **********************************
         proc=0;w=1;kk(:,ga)=0
         do r=1,Q(ga)-1
          z=K(r,ga)
          kk(r,ga)=modulo((a-proc)/w,z)
          proc=proc+kk(r,ga)*w
          w=w*z
         end do
!        **********************************
!        computes the product of A's
!        **********************************
         prodA=1d0
         do r=1,Q(ga)-1
          prodA=prodA*AA(kk(r,ga),r,ga)
         end do
!        **********************************
!        sum over beta begins
!        **********************************
         do b=0,m**(Ne-Q(ga)+1)-1
!         *********************************
!         generates the ii indices
!         *********************************
          proc=0;w=1;ii(:,ga)=0
          do r=1,Q(ga)
           if(nn(r,ga).ne.0) then
            z=m**nn(r,ga)
            ii(r,ga)=modulo((b-proc)/w,z) 
            proc=proc+ii(r,ga)*w
            w=w*z
           end if
          end do
!         *********************************
!         generates relevant indices
!         *********************************
          Ind1=ii(Q(ga),ga)+1
          Ind2=ii(Q(ga),ga)+1
          do r=1,Q(ga)-1
           expo=sum(nn(r+1:Q(ga),ga))+Q(ga)-r
           Ind1=Ind1+ii(r,ga)*m**expo+g1(kk(r,ga),r,ga)*m**(expo-1)
           Ind2=Ind2+ii(r,ga)*m**expo+g2(kk(r,ga),r,ga)*m**(expo-1) 
          end do 
          do u=1,p
           psi(Ind1,u+p)=psi(Ind1,u+p)-prodA*psi(Ind2,u)
          end do
         end do 
        end do 
       end do  
       ${\rm !\$OMP \ end \ parallel \ do}$
      end do
 200  continue
      end
\end{lstlisting} 

\end{document}